\documentclass[letter]{aa} 
\usepackage{graphicx}
\usepackage{natbib,twoopt}
\usepackage{flushend}
\usepackage[colorlinks,linkcolor=blue,urlcolor=blue,citecolor=black]{hyperref} 
\bibpunct{(}{)}{;}{a}{}{,}             
\makeatletter
  \newcommandtwoopt{\citeads}[3][][]{\href{http://ui.adsabs.harvard.edu/abs/#3}%
    {\def\hyper@linkstart##1##2{}%
     \let\hyper@linkend\@empty\citealp[#1][#2]{#3}}}
  \newcommandtwoopt{\citepads}[3][][]{\href{http://ui.adsabs.harvard.edu/abs/#3}%
    {\def\hyper@linkstart##1##2{}%
     \let\hyper@linkend\@empty\citep[#1][#2]{#3}}}
  \newcommandtwoopt{\citetads}[3][][]{\href{http://ui.adsabs.harvard.edu/abs/#3}%
    {\def\hyper@linkstart##1##2{}%
     \let\hyper@linkend\@empty\citet[#1][#2]{#3}}}
  \newcommandtwoopt{\citeyearads}[3][][]%
    {\href{http://ui.adsabs.harvard.edu/abs/#3}
    {\def\hyper@linkstart##1##2{}%
     \let\hyper@linkend\@empty\citeyear[#1][#2]{#3}}}
  \renewcommand*\aa@pageof{, page \thepage{} of \pageref*{LastPage}} 
\makeatother
\usepackage{txfonts}
\begin{document} 

\title{PG\,1610$+$062: a runaway B star challenging classical ejection mechanisms
}
\author{A.~Irrgang\inst{\ref{remeis}}
        \and
        S.~Geier\inst{\ref{potsdam}}
        \and
        U.~Heber\inst{\ref{remeis}}
        \and
        T.~Kupfer\inst{\ref{kavli}}        
        \and
        F.~F\"urst\inst{\ref{caltech},\ref{esac}}
       }
\institute{
Dr.~Karl~Remeis-Observatory \& ECAP, Astronomical Institute, Friedrich-Alexander University Erlangen-Nuremberg (FAU), Sternwartstr.~7, 96049 Bamberg, Germany\\ \email{andreas.irrgang@fau.de}\label{remeis}
\and
Institut f\"ur Physik und Astronomie, Universit\"at Potsdam, Karl-Liebknecht-Str.\ 24/25, 14476 Potsdam, Germany\label{potsdam}
\and
Kavli Institute for Theoretical Physics, University of California, Santa Barbara, CA 93106, USA\label{kavli}
\and
Division of Physics, Mathematics, and Astronomy, California Institute of Technology, Pasadena, CA 91125, USA\label{caltech}
\and
European Space Astronomy Centre (ESA/ESAC), Operations Department, Villanueva de la Canada (Madrid), Spain\label{esac}
}
\date{Received 8 March 2019 / Accepted 10 July 2019}

\abstract{Hypervelocity stars are rare objects, mostly main-sequence (MS) B stars, traveling so fast that they will eventually escape from the Milky Way. Recently, it has been shown that the popular Hills mechanism, in which a binary system is disrupted via a close encounter with the supermassive black hole at the Galactic center, may not be their only ejection mechanism. The analyses of {\it Gaia} data ruled out a Galactic center origin for some of them, and instead indicated that they are extreme disk runaway stars ejected at velocities exceeding the predicted limits of classical scenarios (dynamical ejection from star clusters or binary supernova ejection). We present the discovery of a new extreme disk runaway star, PG\,1610$+$062, which is a slowly pulsating B star bright enough to be studied in detail. A quantitative analysis of spectra taken with ESI at the Keck Observatory revealed that PG\,1610$+$062 is a late B-type MS star of 4--5\,$M_\odot$ with low projected rotational velocity. Abundances (C, N, O, Ne, Mg, Al, Si, S, Ar, and Fe) were derived differentially with respect to the normal B star HD\,137366 and indicate that PG\,1610$+$062 is somewhat metal rich. A kinematic analysis, based on our spectrophotometric distance ($17.3$\,kpc) and on proper motions from {\it Gaia}'s second data release, shows that PG\,1610$+$062 was probably ejected from the Carina-Sagittarius spiral arm at a velocity of $550\pm40$\,km\,s$^{-1}$, which is beyond the classical limits. Accordingly, the star is in the top five of the most extreme MS disk runaway stars and is only the second among the five for which the chemical composition is known.
}


\maketitle
\section{Introduction}\label{section:introduction}
Young stars are expected to be found close to their birthplaces, namely the star-forming regions in the Galactic disk. Finding them far away in the Galactic halo implies that they have been forced to leave their primal environment. Two mechanisms are usually discussed in the literature to explain the presence of these so-called runaway stars (see, e.g., \citeads{2001A&A...365...49H} and references therein). In the binary-supernova scenario \citepads{1961BAN....15..265B} the massive primary star of a binary explodes as a core-collapse supernova and the secondary component is released at almost orbital velocity. In the dynamic scenario \citepads{1967BOTT....4...86P} the runaway stars are formed via gravitational interactions in young and dense stellar clusters, for instance close binary-binary encounters, where the least massive star is usually set free. With typical ejection velocities below a few hundred km\,s$^{-1}$, both of these disk runaway scenarios are by far less powerful than the Hills mechanism \citepads{1988Natur.331..687H}, which describes the disruption of a binary system during a close flyby of the supermassive black hole at the Galactic center (GC). Due to the strong tidal forces, one component is captured while the other is able to leave the site at very high velocity (up to thousands of km\,s$^{-1}$). To highlight their unique origin (and to follow the nomenclature by \citeads{2015AJ....150...77V}), stars stemming from this particular mechanism are referred to as Hills stars in this work. Apart from their formation channels, ejected stars may also be classified according to whether they are gravitationally bound to or unbound from the Milky Way. Stars exceeding their local escape velocity from the Galaxy are commonly called hypervelocity stars (HVSs), the first of which were discovered in 2005 (\citeads{2005ApJ...622L..33B}, \citeads{2005A&A...444L..61H}, \citeads{2005ApJ...634L.181E}). A dedicated spectroscopic survey covering 29\% of the sky revealed 21 candidate HVSs, all of which are late B-type stars that are unbound from the Milky Way if they are main-sequence (MS) stars, and thus at distances of 50--120\,kpc \citepads{2014ApJ...787...89B}. Until recently, the Hills mechanism was widely assumed to be the only ejection scenario that is capable of producing MS HVSs \citepads{2015ARA&A..53...15B}. However, high-precision astrometry from {\it Gaia}'s second data release (DR; \citeads{2018A&A...616A...1G}) shows that some of the candidate HVSs no longer qualify as Hills stars because the GC can be most likely ruled out as their spatial origin \citepads{2018A&A...620A..48I}. Because the ejection velocities of those dismissed Hills stars are higher than the upper limits for the two ``classical'' disk ejection scenarios mentioned above, a powerful yet neglected or unknown mechanism (e.g., dynamical interactions with massive stars or intermediate-mass black holes) must be at work \citepads{2018A&A...620A..48I}. To gain deeper insights, more stars ejected by this mechanism have to be studied in detail. Here, we investigate PG\,1610$+$062, a blue star at high Galactic latitude ($b=+37.80\degr$). It was first discovered during the Palomar-Green survey \citepads{1986ApJS...61..305G} where it was classified as a horizontal branch B star. Apart from a re-classification as MS B-type star by \citetads{2015A&A...577A..26G}, no attempt has been made since then to study this object in more detail. The star attracted our attention in the course of the MUCHFUSS project \citepads{2011A&A...530A..28G} because a set of low-resolution spectra indicated that its radial velocity might be variable \citepads{2015A&A...577A..26G}. Unlike the faint stars of the HVS sample, which have visual magnitudes between 17.5 and 20\,mag \citepads{2014ApJ...787...89B}, PG\,1610$+$062 is bright enough ($V=15.6$\,mag) for a high-precision quantitative spectroscopic (Sect.~\ref{section:spectroscopy}) and photometric (Sect.~\ref{section:photometry}) analysis. A kinematic investigation (Sect.~\ref{section:kinematics}) yields an ejection velocity of $550\pm40$\,km\,s$^{-1}$, granting PG\,1610$+$062 a place in the top five of the most extreme disk runaway MS stars known to date (Sect.~\ref{section:summary}).
\section{Spectroscopic analysis}\label{section:spectroscopy}
\begin{table*}
\centering
\footnotesize
\setlength{\tabcolsep}{0.093cm}
\renewcommand{\arraystretch}{1.1}
\caption{\label{table:atmospheric_parameters} Atmospheric parameters and abundances of the two program stars.}
\begin{tabular}{lrrrrrrrrrrrrrrrrrr}
\hline\hline
Object & $T_{\textnormal{eff}}$ & $\log(g)$ & $\varv_{\textnormal{rad}}$ & $\varv\sin(i)$ & $\zeta$ & $\xi$ & & \multicolumn{11}{c}{$\log(n(x))$} \\ 
\cline{4-7} \cline{9-19}
& (K) & (cgs) & \multicolumn{4}{c}{(km\,s$^{-1}$)} & & He & C & N & O & Ne & Mg & Al & Si & S & Ar & Fe\\
\hline
PG\,1610$+$062 & $14\,800$ & $4.054$ & $157.4$\tablefootmark{a} & $15.5$ & $0.0$ & $2.40$ &   & $-0.92$ & $-3.34$ & $-3.86$ & $-3.09$ & $-3.87$ & $-4.51$ & $-5.67$ & $-4.24$ & $-4.66$ & $-5.32$ & $-4.36$ \\ Stat. & $^{+80}_{-80}$ & $^{+0.022}_{-0.023}$ & $7.7$\tablefootmark{a} & $^{+1.4}_{-1.5}$ & $^{+15.0}_{-\phantom{0}0.0}$ & $^{+0.27}_{-0.26}$ &   & $^{+0.04}_{-0.03}$ & $^{+0.07}_{-0.07}$ & $^{+0.12}_{-0.12}$ & $^{+0.05}_{-0.05}$ & $^{+0.04}_{-0.04}$ & $^{+0.06}_{-0.06}$ & $^{+0.08}_{-0.07}$ & $^{+0.06}_{-0.07}$ & $^{+0.05}_{-0.04}$ & $^{+0.22}_{-0.33}$ & $^{+0.06}_{-0.05}$ \\ Sys. & $^{+300}_{-300}$ & $^{+0.100}_{-0.100}$ & \ldots\tablefootmark{a} & $^{+0.6}_{-0.5}$ & $^{+0.1}_{-0.0}$ & $^{+0.16}_{-0.17}$ &   & $^{+0.12}_{-0.14}$ & $^{+0.08}_{-0.07}$ & $^{+0.04}_{-0.04}$ & $^{+0.06}_{-0.06}$ & $^{+0.05}_{-0.04}$ & $^{+0.07}_{-0.07}$ & $^{+0.05}_{-0.04}$ & $^{+0.07}_{-0.08}$ & $^{+0.06}_{-0.05}$ & $^{+0.10}_{-0.06}$ & $^{+0.13}_{-0.14}$ \\
    HD\,137366 & $14\,930$ & $3.803$ &                  $-15.5$ & $11.3$ & $2.5$ & $1.91$ &   & $-0.97$ & $-3.52$ & $-4.15$ & $-3.16$ & $-3.99$ & $-4.66$ & $-5.83$ & $-4.35$ & $-4.87$ & $-5.55$ & $-4.51$ \\ Stat. & $^{+10}_{-20}$ & $^{+0.002}_{-0.002}$ &       $^{+0.1}_{-0.1}$ & $^{+0.1}_{-0.1}$ &             $^{+0.4}_{-0.6}$ & $^{+0.07}_{-0.07}$ &   & $^{+0.01}_{-0.01}$ & $^{+0.02}_{-0.02}$ & $^{+0.02}_{-0.02}$ & $^{+0.02}_{-0.01}$ & $^{+0.01}_{-0.02}$ & $^{+0.02}_{-0.02}$ & $^{+0.02}_{-0.03}$ & $^{+0.02}_{-0.02}$ & $^{+0.01}_{-0.01}$ & $^{+0.04}_{-0.04}$ & $^{+0.01}_{-0.02}$ \\ Sys. & $^{+300}_{-300}$ & $^{+0.100}_{-0.100}$ &        $^{+0.1}_{-0.1}$ & $^{+0.2}_{-0.3}$ & $^{+0.1}_{-0.1}$ & $^{+0.42}_{-0.66}$ &   & $^{+0.12}_{-0.13}$ & $^{+0.09}_{-0.09}$ & $^{+0.06}_{-0.09}$ & $^{+0.02}_{-0.03}$ & $^{+0.03}_{-0.03}$ & $^{+0.05}_{-0.06}$ & $^{+0.05}_{-0.06}$ & $^{+0.05}_{-0.05}$ & $^{+0.05}_{-0.04}$ & $^{+0.05}_{-0.07}$ & $^{+0.07}_{-0.11}$ \\

\hline
\end{tabular}
\tablefoot{The abundance $n(x)$ is given as fractional particle number of species $x$ with respect to all elements. Statistical uncertainties (\textit{``Stat.''}) correspond to $\Delta \chi^2 = 6.63$ and are 99\% confidence limits. Systematic uncertainties (\textit{``Sys.''}) cover only the effects induced by additional variations of $2\%$ in $T_{\textnormal{eff}}$ and $0.1$ in $\log(g)$ and are formally taken to be 99\% confidence limits (see \citeads{2014A&A...565A..63I} for details). \tablefoottext{a}{The radial velocity of PG\,1610$+$062 is given here as the average over the four results from the ESI spectra. This value is consistent with radial velocities measured in low-resolution spectra taken at more than a dozen different epochs.}}
\end{table*}
\begin{table*}
\centering
\footnotesize
\setlength{\tabcolsep}{0.116cm}
\renewcommand{\arraystretch}{1.1}
\caption{\label{table:stellar_parameters} Stellar parameters, spectrophotometric distances, color excesses, and photospheric mass fractions of the two program stars.}
\begin{tabular}{lcrrcrrcrrcrrcrrcrrcrrcrrcrr}
\hline\hline
Object & & \multicolumn{2}{c}{$M$} & & \multicolumn{2}{c}{$\tau$} & & \multicolumn{2}{c}{$\log(L/L_{\sun})$} & & \multicolumn{2}{c}{$R_{\star}$} & & \multicolumn{2}{c}{$d$} & & \multicolumn{2}{c}{$E(B-V)$} & & \multicolumn{2}{c}{$X$} & & \multicolumn{2}{c}{$Y$} & & \multicolumn{2}{c}{$Z$}\\ 
\cline{3-4} \cline{6-7} \cline{9-10} \cline{12-13} \cline{15-16} \cline{18-19} \cline{21-22} \cline{24-25} \cline{27-28}
& & \multicolumn{2}{c}{$(M_{\sun})$} & & \multicolumn{2}{c}{(Myr)} & & & & & \multicolumn{2}{c}{$(R_{\sun})$} & & \multicolumn{2}{c}{(pc)} & & \multicolumn{2}{c}{(mag)} & & & & & & & & & \\
\hline
PG\,1610$+$062 &   & $4.4$ & $^{+0.3}_{-0.2}$ &   & $83$ &           $^{+22}_{-24}$ &   & $2.66$ & $^{+0.11}_{-0.10}$ &   & $3.3$ & $^{+0.5}_{-0.5}$ &   & $17\,300$ & $^{+2910}_{-2480}$ &   & $0.024$ & $^{+0.025}_{-0.024}$ &   & $0.635$ & $^{+0.069}_{-0.074}$ &   & $0.344$ & $^{+0.077}_{-0.071}$ &   & $0.021$ & $^{+0.003}_{-0.003}$ \\
    HD\,137366 &   & $5.1$ & $^{+0.3}_{-0.3}$ &   & $83$ & $^{+14}_{-\phantom{0}8}$ &   & $2.99$ & $^{+0.11}_{-0.09}$ &   & $4.7$ & $^{+0.8}_{-0.7}$ &   &     $360$ &     $^{+60}_{-50}$ &   & $0.056$ & $^{+0.020}_{-0.019}$ &   & $0.667$ & $^{+0.063}_{-0.067}$ &   & $0.316$ & $^{+0.069}_{-0.063}$ &   & $0.016$ & $^{+0.002}_{-0.001}$ \\

\hline
\end{tabular}
\tablefoot{Except for the distance $d$, for which photometric uncertainties are included in the error budget, uncertainties cover only the effects induced by variations of $2\%$ in $T_{\textnormal{eff}}$ and $0.1$ in $\log(g)$ and are formally taken to be 99\% confidence limits (see \citeads{2014A&A...565A..63I} for details).} 
\end{table*}
\begin{figure}
\centering
\includegraphics[width=0.49\textwidth]{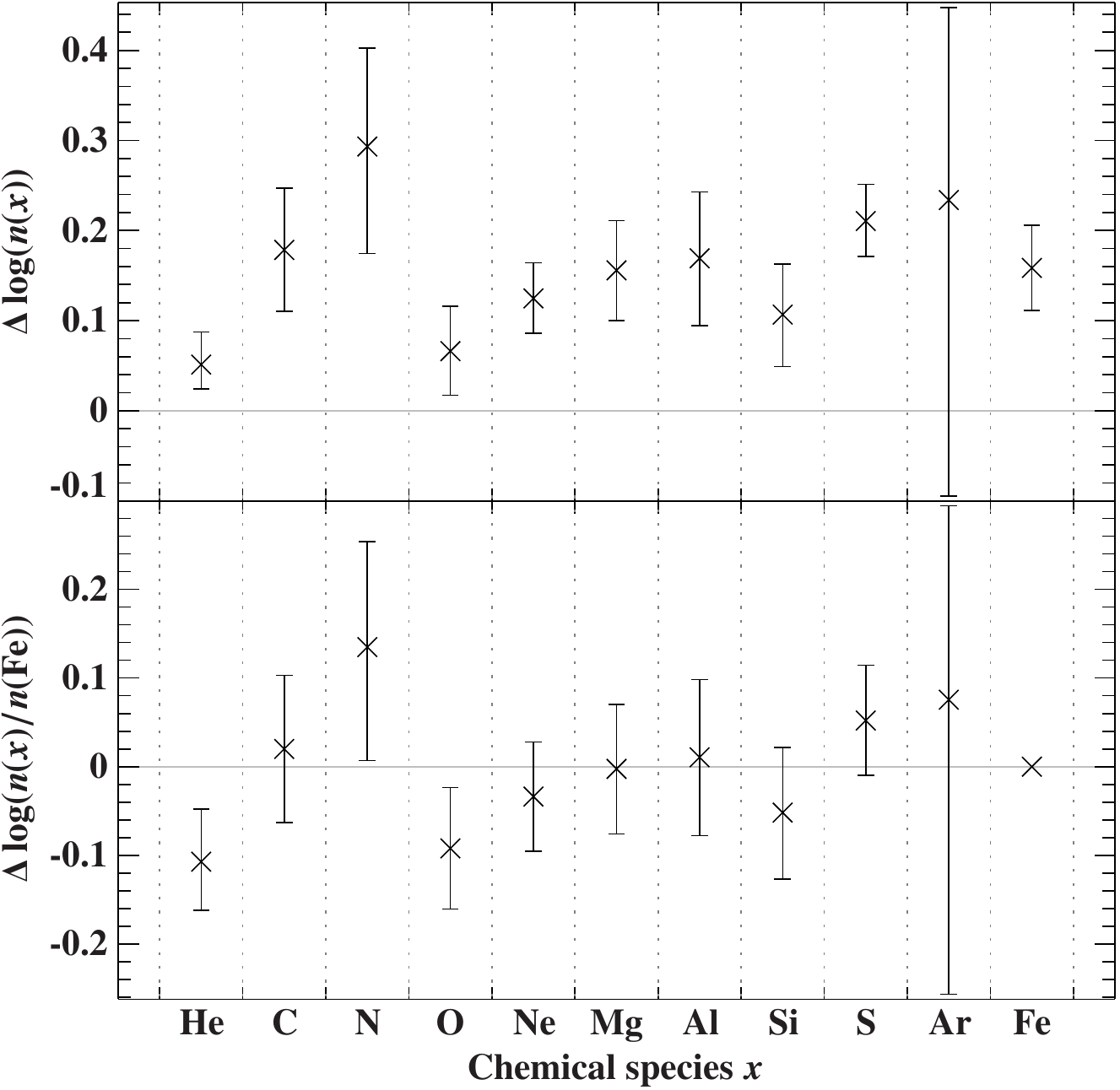}
\caption{Differential abundance pattern (\textit{top}) and element-to-iron abundance ratios (\textit{bottom}) of PG\,1610$+$062 with respect to the solar neighborhood reference star HD\,137366. The error bars are the square roots of the quadratic sums of the statistical uncertainties given in Table~\ref{table:atmospheric_parameters}, and thus represent 99\% confidence intervals.}
\label{fig:abundance_pattern}
\end{figure}
The spectroscopic analysis is based on four medium-resolution ($R \approx 8000$) spectra of decent individual signal-to-noise ratio ($\textnormal{S/N} \approx 60-110$ in the visual) obtained with the ESI spectrograph \citepads{2002PASP..114..851S} mounted at the Keck~II telescope and based on low-resolution spectra taken with the SDSS spectrograph (\citeads{2013AJ....146...32S}, $R \approx 1500-2500$, $\textnormal{S/N} \approx 102$ in the visual) and the {\sc Twin} spectrograph\footnote{\url{http://www.caha.es/pedraz/Twin/}} mounted at the 3.5\,m telescope at Calar Alto observatory ($R \approx 1500-2500$, $\textnormal{S/N} \approx 160$ in the visual, coadded from 14 individual spectra).

The quantitative analysis strategy and the applied models are explained in detail in \citetads{2014A&A...565A..63I}. In short, a simultaneous fit of all spectra over their entire spectral range is performed to constrain all parameters (i.e., atmospheric parameters and chemical abundances) at the same time. The underlying synthetic spectra are based on the hybrid approach, where the structure of the atmosphere is computed in local thermodynamic equilibrium (LTE) with {\sc Atlas12} \citepads{1996ASPC..108..160K}. Departures from LTE are then accounted for by applying updated versions of {\sc Detail} and {\sc Surface} (\citeads{1981PhDT.......113G}; \citealt{detailsurface2}). The {\sc Detail} code computes population numbers in non-LTE by numerically solving the coupled radiative transfer and statistical equilibrium equations. The {\sc Surface} code uses the resulting departure coefficients and more detailed line-broadening data to compute the final synthetic spectrum. All three codes have been recently updated to allow for level dissolution of hydrogen --~following the description by \citetads{1994A&A...282..151H} and using line broadening tables by \citetads{2009ApJ...696.1755T}~-- and non-LTE feedback on the atmospheric structure \citepads{2018A&A...615L...5I}.

In order to minimize systematic uncertainties, we carried out a differential abundance analysis with respect to the B-type star HD\,137366, for which a high-quality spectrum ($R \approx 48\,000$, $\textnormal{S/N} \approx 470$ in the visual) taken with {\sc Feros} \citepads{1999Msngr..95....8K} is available. This particular object was chosen because, on the one hand, it is nearby and thus a representative of B-type stars in the solar neighborhood and, on the other hand, it is almost a spectroscopic twin of PG\,1610$+$062 (see Figs.~\ref{fig:spectra_1}--\ref{fig:spectra_9}) making it an ideal target for a differential abundance study.

The results of the spectroscopic analyses are summarized in Table~\ref{table:atmospheric_parameters}. Both objects are late B-type stars with slow projected rotation ($\varv\sin(i) \sim 15$\,km\,s$^{-1}$), i.e., they exhibit very sharp metal lines. While the abundance pattern of HD\,137366 is very similar to those of other young B-type stars in the solar neighborhood (cf.~\citeads{2012A&A...539A.143N}), there is a uniform enrichment in the elemental abundances of PG\,1610$+$062 (Fig.~\ref{fig:abundance_pattern}) which indicates a higher baseline metallicity. At least to some extent, this is expected because the star originates $\sim 1.8$\,kpc closer to the GC than the Sun (see Sect.~\ref{section:kinematics}) so that Galactic abundance gradients (see, e.g., \citeads{2012A&A...539A.143N} and references therein) come into play. Stellar parameters (Table~\ref{table:stellar_parameters}) are based on comparing the stars' positions in a $(T_{\textnormal{eff}},\log(g))$ diagram to single-star evolutionary tracks by \citetads{2013A&A...553A..24G}. The two objects are consistent with being young ($\sim 83$\,Myr), massive ($\sim 4$--$5$\,$M_\sun$) MS stars. In principle, the derived values for $T_{\textnormal{eff}}$, $\log(g)$, and $\varv\sin(i)$ also fit those of blue horizontal branch stars. This option, however, is very unlikely because the abundance patterns of those evolved objects are strongly altered by diffusion processes, an effect that is not observed here.
\section{Photometric analysis}\label{section:photometry}
\subsection{Spectral energy distribution}
\begin{table}
\centering
\footnotesize
\renewcommand{\arraystretch}{1.1}
\caption{Stellar parameters derived from photometry.}
\label{table:photometry}
\begin{tabular}{lr}
\hline\hline
Parameter & Value \\
\hline
\multicolumn{2}{l}{PG\,1610$+$062:} \\
Effective temperature $T_{\textnormal{eff}}$ & $14\,800^{+2500}_{-1100}$\,K \\
Surface gravity $\log (g\,(\textnormal{cm}\,\textnormal{s}^{-2}))$ & $3.2^{+1.7}_{-1.2}$ \\
Angular diameter $\Theta$ & $\left(8.6\pm0.5\right)\times10^{-12}$\,rad \\
Color excess $E(B-V)$ & $\le 0.09$\,mag \\
\hline
\multicolumn{2}{l}{HD\,137366:} \\
Effective temperature $T_{\textnormal{eff}}$ & $15\,000\pm900$\,K \\
Surface gravity $\log (g\,(\textnormal{cm}\,\textnormal{s}^{-2}))$ & $3.6\pm0.9$ \\
Angular diameter $\Theta$ & $(5.91\pm0.19)\times10^{-10}$\,rad \\
Color excess $E(B-V)$ & $0.056\pm0.018$\,mag \\
\hline
\end{tabular}
\tablefoot{The given uncertainties are single-parameter 99\% confidence intervals based on $\chi^2$ statistics.}
\end{table}
Spectral energy distributions (see Fig.~\ref{fig:photometry}) were also investigated in order to cross-check atmospheric parameters and to derive spectrophotometric distances. Table~\ref{table:photometry} lists the parameters derived from fitting {\sc Atlas12} models to the available photometric measurements. For both targets, spectroscopic and photometric results are consistent with each other, with almost identical effective temperatures. The spectrophotometric distances $d$ given in Table~\ref{table:stellar_parameters} are based on the corresponding stellar radii $R_\star$ and on the angular diameters $\Theta = 2 R_\star / d$ from Table~\ref{table:photometry}. For the nearby reference star HD\,137366, the parallax measurement from {\it Gaia} DR2 is highly significant ($\varpi = 2.8014 \pm 0.0566$\,mas) and can thus be exploited as a consistency check. The agreement between parallactic ($1/\varpi = 357\pm8$\,pc) and spectrophotometric distance (see Table~\ref{table:stellar_parameters}) is perfect, validating its MS nature and showing that our spectrophotometric distance estimates are trustworthy. This is important for PG\,1610$+$062, which is quite distant ($d=17.30^{+2.91}_{-2.48}$\,kpc), and hence has a highly uncertain {\it Gaia} parallax ($\varpi = 0.0143 \pm 0.0520$\,mas).
\subsection{Light curve}
PG\,1610$+$062 lies right inside the instability domain of slowly pulsating B (SPB) stars (see, e.g., \citeads{2016MNRAS.455L..67M}) and is thus expected to pulsate if it is a MS star. The ATLAS variable star catalog \citepads{2018AJ....156..241H} indeed classifies it as a candidate variable star. Because this classification is based on a purely automated procedure, we decided to reanalyze the ATLAS data to test the robustness of the results. The outcome of this exercise, which is presented in the Appendix, confirms the oscillation period reported by \citetads{2018AJ....156..241H} of $4.336721$\,days. Moreover, it shows that PG\,1610$+$062 exhibits oscillation properties (see Table~\ref{table:oscillation_params}) that are characteristic of SPB stars (see, e.g., \citeads{2015pust.book.....C} and references therein). Another typical feature of SPB stars are temporal distortions of the line profiles (see, e.g., \citeads{2016A&A...591L...6I}). Spectra with very high spectral resolution and S/N are required to resolve them, which unfortunately are not available. At the limited quality of our spectra, the variations may only lead to small changes in the radial velocity on the order of a few km\,s$^{-1}$, which we and \citetads{2015A&A...577A..26G} indeed observed. We conclude that PG\,1610$+$062 is an SPB star, which supports our classification as a MS star.
\section{Kinematic analysis}\label{section:kinematics}
\begin{figure}
\centering
\includegraphics[width=0.49\textwidth]{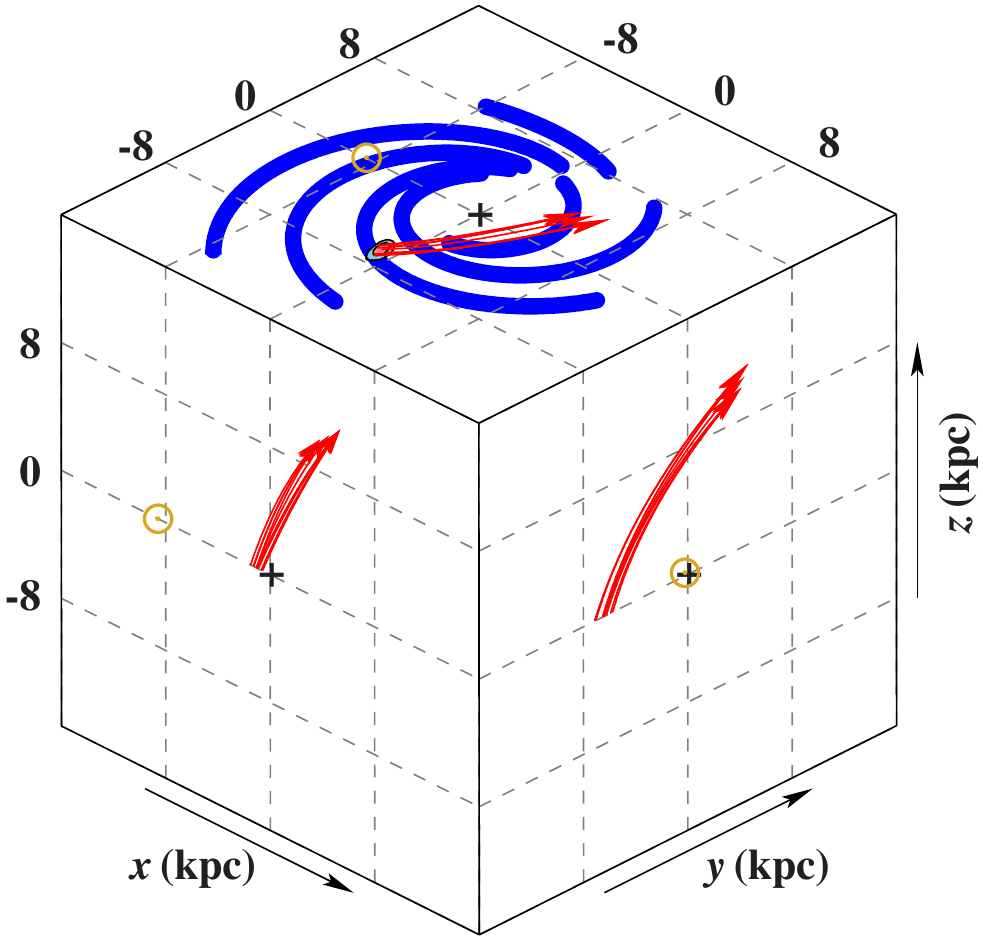}
\caption{Three-dimensional orbit of PG\,1610$+$062 in a Galactic Cartesian coordinate system in which the $z$-axis points to the Galactic north pole. The nine trajectories (red lines; arrows indicate the star's current position) are computed in Model~I of \citetads{2013A&A...549A.137I} and illustrate the effects of uncertainties in the distance, proper motions, and radial velocity. Orbits were computed back in time until they reached the Galactic plane. The small black rimmed, shaded areas are $1\sigma$ (red) and $2\sigma$ (light blue) contours for the intersection of the Galactic plane. The thick blue solid lines schematically represent the loci of the spiral arms 41\,Myr ago based on the polynomial logarithmic arm model of \citetads{2014A&A...569A.125H} and the Galactic rotation curve of Model~I of \citetads{2013A&A...549A.137I}. The current positions of the Sun and the GC are marked by a yellow circled dot ($\odot$) and a black plus sign ($+$), respectively. The orbit is characteristic of a disk runaway star.
}\label{fig:orbits_PG1610}
\end{figure}
\begin{table*}
\centering
\footnotesize
\setlength{\tabcolsep}{0.135cm}
\caption{\label{table:kinematic_parameters} Kinematic parameters of PG\,1610$+$062 for the three different Galactic mass models of \citetads{2013A&A...549A.137I}.}
\begin{tabular}{lrrrrrrrrrrrrrrrrrrrrrr}
\hline\hline
 & $x$ & $y$ & $z$ & $r$ & & $\varv_x$ & $\varv_y$ & $\varv_z$ & $\varv_{\textnormal{Grf}}$ & $\varv_{\textnormal{Grf}}-\varv_{\textnormal{esc}}$ & $P_{\textnormal{b}}$ & $x_{\textnormal{d}}$ & $y_{\textnormal{d}}$ & $z_{\textnormal{d}}$ & $r_{\textnormal{d}}$ & & $\varv_{x\textnormal{,d}}$ & $\varv_{y\textnormal{,d}}$ & $\varv_{z\textnormal{,d}}$ & $\varv_{\textnormal{Grf,d}}$ & $\varv_{\textnormal{ej}}$ & $\tau_{\textnormal{flight}}$ \\
\cline{2-5} \cline{7-11} \cline{13-16} \cline{18-22}
& \multicolumn{4}{c}{(kpc)} & & \multicolumn{5}{c}{(km\,s$^{-1}$)} & (\%) & \multicolumn{4}{c}{(kpc)} & & \multicolumn{5}{c}{(km\,s$^{-1}$)} & (Myr) \\
\hline \hline
Model~I & $4.6$ & $4.4$ & $10.6$ & $12.4$ &   & $100$ & $270$ & $140$ & $320$ & $-250$ & $100$ & $-1.1$ & $-6.4$ & $-0.0$ & $6.5$ &   & $150$ & $160$ & $370$ & $430$ & $550$ & $41$ \\ Stat. & $^{+2.2}_{-1.9}$ & $^{+0.8}_{-0.7}$ & $^{+1.8}_{-1.6}$ & $^{+2.7}_{-2.2}$ &   & $^{+20}_{-10}$ & $^{+20}_{-20}$ & $^{+20}_{-10}$ & $^{+20}_{-10}$ & $^{+30}_{-20}$ & \ldots & $^{+0.9}_{-0.6}$ & $^{+1.1}_{-1.5}$ & $^{+0.1}_{-0.1}$ & $^{+1.4}_{-1.0}$ &   & $^{+40}_{-50}$ & $^{+20}_{-20}$ & $^{+30}_{-20}$ & $^{+20}_{-10}$ & $^{+40}_{-40}$ & $^{+8}_{-7}$ \\

Model~II & $4.6$ & $4.4$ & $10.6$ & $12.4$ &   & $100$ & $270$ & $140$ & $320$ & $-200$ & $100$ & $-1.0$ & $-6.3$ & $0.0$ & $6.4$ &   & $150$ & $160$ & $370$ & $430$ & $550$ & $41$ \\ Stat. & $^{+2.2}_{-1.9}$ & $^{+0.8}_{-0.7}$ & $^{+1.8}_{-1.6}$ & $^{+2.7}_{-2.2}$ &   & $^{+20}_{-10}$ & $^{+20}_{-20}$ & $^{+20}_{-10}$ & $^{+20}_{-10}$ & $^{+20}_{-30}$ & \ldots & $^{+0.8}_{-0.7}$ & $^{+1.1}_{-1.5}$ & $^{+0.1}_{-0.1}$ & $^{+1.5}_{-1.0}$ &   & $^{+40}_{-40}$ & $^{+20}_{-20}$ & $^{+30}_{-20}$ & $^{+20}_{-10}$ & $^{+40}_{-30}$ & $^{+8}_{-7}$ \\

Model~III & $4.6$ & $4.4$ & $10.6$ & $12.4$ &   & $100$ & $270$ & $140$ & $320$ & $-460$ & $100$ & $-1.0$ & $-6.5$ & $-0.0$ & $6.6$ &   & $150$ & $160$ & $370$ & $430$ & $550$ & $41$ \\ Stat. & $^{+2.2}_{-1.9}$ & $^{+0.8}_{-0.7}$ & $^{+1.8}_{-1.6}$ & $^{+2.7}_{-2.2}$ &   & $^{+20}_{-10}$ & $^{+20}_{-20}$ & $^{+20}_{-10}$ & $^{+20}_{-10}$ & $^{+30}_{-20}$ & \ldots & $^{+0.9}_{-0.7}$ & $^{+1.1}_{-1.4}$ & $^{+0.1}_{-0.1}$ & $^{+1.3}_{-1.0}$ &   & $^{+40}_{-40}$ & $^{+20}_{-10}$ & $^{+20}_{-30}$ & $^{+20}_{-20}$ & $^{+40}_{-40}$ & $^{+8}_{-6}$ \\

\hline
\end{tabular}
\tablefoot{Results and statistical uncertainties (\textit{``Stat.''}) are given as median values and 99\% confidence limits which are derived via a Monte Carlo simulation. The Galactic coordinate system is introduced in Fig.~\ref{fig:orbits_PG1610}. Disk-crossing quantities are labeled by the subscript ``d''. The Galactic rest-frame velocity $\varv_{\textnormal{Grf}}=(\varv_x^2+\varv_y^2+\varv_z^2)^{1/2}$, the local Galactic escape velocity $\varv_{\textnormal{esc}}$, the Galactocentric radius $r=(x^2+y^2+z^2)^{1/2}$, the ejection velocity $\varv_{\textnormal{ej}}$ (defined as the Galactic rest-frame velocity relative to the rotating Galactic disk), and the flight time $\tau_{\textnormal{flight}}$ are listed in addition to Cartesian positions and velocities. The probability $P_{\textnormal{b}}$ is the fraction of Monte Carlo runs for which the star is bound to the Milky Way.}
\end{table*}
To investigate the origin of PG\,1610$+$062, a detailed kinematic investigation was carried out (i.e., the star's trajectory was traced back to the Galactic disk; see Fig.~\ref{fig:orbits_PG1610}). Only the most recent disk-crossing event is considered here because all the others occurred too far in the past to be compatible with the lifetime of the star. Systematic uncertainties were estimated by applying three different models for the gravitational potential of the Milky Way, all of which are axisymmetric three-component models with identical mathematical forms for the bulge and disk, but with their own parameter values and varying expressions for the dark matter halo component (see \citeads{2013A&A...549A.137I} for details). Because the Galactocentric radii traveled by the program star (roughly between $12.4$--$6.5$\,kpc) lie in a regime where the halo component is the dominating acceleration force (see, e.g., Fig.~1 in \citeads{2008ApJ...680..312K}), these three models are ideally suited to assess systematic uncertainties, which turned out to be completely negligible. A comparison with recent analyses of the motion of globular clusters, satellite galaxies, and extreme velocity stars shows that the models' mass distributions and local escape velocities are consistent with results from {\it Gaia} DR2 astrometry \citepads{2018A&A...620A..48I}. Statistical uncertainties in the spectrophotometric distance, radial velocity, and proper motions from {\it Gaia} DR2 (\citeads{2018A&A...616A...2L}; {\it Gaia}\,DR2\,4450123955938796160, $\mu_\alpha \cos\delta = -0.616 \pm 0.076$\,mas\,yr$^{-1}$, $\mu_\delta = 0.176 \pm 0.042$\,mas\,yr$^{-1}$) were propagated via a Monte Carlo procedure with $100\,000$ runs that simultaneously and independently varies the individual parameters assuming Gaussian distributions for each parameter, while also accounting for asymmetric error bars and the correlation ($0.5031$) between the two proper motion components. The outcome of the kinematic analysis is summarized in Table~\ref{table:kinematic_parameters} and is perfectly consistent with a Galactic disk runaway scenario. PG\,1610$+$062 was shot into the halo $\sim 41$\,Myr ago from a region with a Galactocentric radius of $\sim 6.5$\,kpc, which possibly coincided with the location of the now nearby Carina-Sagittarius spiral arm. Despite its huge ejection velocity of $550\pm40$\,km\,s$^{-1}$, it is still gravitationally bound to the Milky Way because the ejection vector was somewhat opposite to Galactic rotation. 

In contrast, the kinematic properties of HD\,137366 are typical of thin-disk stars in the solar neighborhood (see Fig.~\ref{fig:orbits_HD137366}).
\section{Summary and discussion}\label{section:summary}
\begin{table}
\centering
\renewcommand{\arraystretch}{1.07}
\caption{Ejection velocities (relative to the rotating Galactic disk) and heliocentric distances of candidate MS stars that were possibly ejected from the Galactic disk beyond the velocity limit of classical mechanisms.}
\label{table:ejection_velocities}
\footnotesize
\begin{tabular}{llll}
\hline\hline
Star & $\varv_\textnormal{ej}$\,(km\,s$^{-1}$) & Distance\,(kpc) & Reference \\
\hline
HVS\,5                  & $640^{+50}_{-40}$   & $31.2^{+3.2}_{-2.5}$ & (1), (2) \\
B711                    & $600^{+90}_{-50}$   & $28.5^{+3.1}_{-2.2}$ & (1), (2) \\
B434                    & $590\pm20$          & $40.5^{+4.7}_{-3.7}$ & (1), (2) \\
LAMOST-HVS1             & $568^{+19}_{-17}$   & $19.1^{+5.1}_{-3.8}$ & (3) \\
\textbf{PG\,1610$+$062} & $\mathbf{550\pm20}$ & $\mathbf{17.3^{+1.2}_{-1.0}}$ & \textbf{This work} \\
HVS\,7                  & $530\pm30$          & $48.2^{+4.3}_{-3.7}$ & (1), (2) \\
HVS\,12                 & $510^{+40}_{-30}$   & $51.7^{+9.0}_{-6.1}$ & (1), (2) \\
LAMOST-HVS4             & $480^{+13}_{-10}$   & $27.9\pm1.5$         & (4)\tablefootmark{a} \\
EC\,19596$-$5356        & $475^{+74}_{-83}$   & $13.81^{+4.80}_{-3.63}$ & (5) \\
HIP\,56322              & $471^{+189}_{-\phantom{0}99}$ & \phantom{0}$6.09^{+3.17}_{-1.92}$ & (5) \\
HIP\,105912             & $457^{+130}_{-133}$ & \phantom{0}$4.17^{+1.70}_{-1.14}$ & (5) \\
HVS\,8                  & $450^{+40}_{-30}$   & $37.2^{+4.4}_{-3.6}$ & (1), (2) \\
B733                    & $450\pm10$          & \phantom{0}$9.9^{+0.7}_{-0.9}$ & (1), (2) \\
BD\,-2\,3766            & $425^{+151}_{-109}$ & \phantom{0}$4.22^{+1.50}_{-1.10}$ & (5) \\
B485                    & $420^{+20}_{-10}$   & $33.3^{+3.7}_{-1.7}$ & (1), (2) \\
PHL\,346                & $418^{+49}_{-47}$   & \phantom{0}$8.55^{+1.61}_{-1.33}$ & (5) \\
PB\,5418                & $415^{+141}_{-100}$ & \phantom{0}$6.09^{+2.03}_{-1.49}$ & (5) \\
PG\,1332$+$137          & $413^{+38}_{-77}$   & \phantom{0}$6.54^{+2.13}_{-1.70}$ & (5) \\
HIP\,114569             & $408^{+89}_{-71}$   & \phantom{0}$1.60^{+0.40}_{-0.31}$ & (5) \\
PHL\,2018               & $399^{+68}_{-66}$   & \phantom{0}$6.93^{+2.39}_{-1.77}$ & (5) \\
PG\,1209$+$263          & $390^{+293}_{-100}$ & $30.93^{+7.94}_{-7.28}$ & (5) \\
HD\,271791              & $390^{+70}_{-30}$   & $21\pm4$             & This work\tablefootmark{b} \\
PG\,0914$+$001          & $369^{+240}_{-157}$ & $20.62^{+6.74}_{-5.28}$ & (5) \\
\hline
\end{tabular}
\tablefoot{The given uncertainties are $1\sigma$ errors. \tablefoottext{a}{Assuming a MS nature, \citetads{2018AJ....156...87L} give a Galactic rest-frame velocity at a disk intersection of $697\pm12$\,km\,s$^{-1}$, which transforms to the given ejection velocity.} \tablefoottext{b}{Based on the distance and radial velocity from \citetads{2008A&A...483L..21H} and on proper motions from {\it Gaia} DR2.}}
\tablebib{
(1)~\citetads{2018A&A...620A..48I}; (2)~\citetads{2018A&A...615L...5I}; (3)~\citetads{2018arXiv181002029H}; (4)~\citetads{2018AJ....156...87L}; (5)~\citetads{2011MNRAS.411.2596S}.
}
\end{table}
With the release of {\it Gaia} DR2, there is growing observational evidence that MS stars can be accelerated to beyond their local Galactic escape velocity from within the Galactic disk, i.e., without the involvement of the supermassive black hole at the GC (\citeads{2018AJ....156...87L}, \citeads{2018A&A...620A..48I}, \citeads{2018arXiv181002029H}). While the first of these unbound disk runaway stars (which are sometimes referred to as hyper-runaway stars; see \citeads{2008ApJ...684L.103P}), HD\,271791, could still be explained in the framework of the ``classical'' disk ejection scenarios outlined in Sect.~\ref{section:introduction}, namely via an extreme case of the supernova mechanism with additional boost by Galactic rotation \citepads{2008ApJ...684L.103P}, this is not the case for most of the other unbound disk runaway candidates because their intrinsic ejection velocities (see Table~\ref{table:ejection_velocities}) exceed the respective upper limits of $\sim 400$\,km\,s$^{-1}$ (see \citeads{2018A&A...620A..48I} for an extensive discussion on the upper limits). Close encounters with very massive stars or intermediate-mass black holes offer, in principle, a straightforward explanation (see, e.g., \citeads{2018A&A...620A..48I} and references therein). However, the rates at which those strong dynamical interactions may occur are not well constrained because the actual number of massive perturbers and the conditions in their host clusters are uncertain (see, e.g., \citeads{2018arXiv181002029H}). With only a few objects known so far, it is still crucial to increase the sample of stars ejected by this powerful mechanism in order to provide tighter observational constraints on the theory. Here, we present the discovery of a new member of this tiny group, PG\,1610$+$062. Owing to the unprecedented precision of proper motions from {\it Gaia} DR2 and that this star is relatively close ($\sim 17.3$\,kpc) compared to many other extreme velocity stars, it is possible to study it in great detail. Our spectroscopic, photometric, and kinematic analyses suggest that PG\,1610$+$062 is a young B-type MS star originating from a relatively small area close to the Carina-Sagittarius spiral arm, which today is not too far away from the Sun. Although it is not gravitationally unbound from the Milky Way, it is in the top five of the most extreme MS disk runaway stars (see Table~\ref{table:ejection_velocities}) and is, after LAMOST-HVS1, only the second of these five for which the chemical composition is known.
\begin{acknowledgements}
We thank John E.\ Davis for the development of the {\sc slxfig} module used to prepare the figures in this paper. 
Based on observations made with ESO Telescopes at the La Silla Paranal Observatory under programme ID 091.C-0713(A). 
Based on data from the CAHA Archive at CAB (INTA-CSIC). 
Some of the data presented herein were obtained at the W.M.\ Keck Observatory, which is operated as a scientific partnership among the California Institute of Technology, the University of California and the National Aeronautics and Space Administration. The Observatory was made possible by the generous financial support of the W.M.\ Keck Foundation. The authors wish to recognize and acknowledge the very significant cultural role and reverence that the summit of Mauna Kea has always had within the indigenous Hawaiian community. We are most fortunate to have the opportunity to conduct observations from this mountain. 
This work has made use of data from the European Space Agency (ESA)
mission {\it Gaia} (\url{https://www.cosmos.esa.int/gaia}), processed by
the {\it Gaia} Data Processing and Analysis Consortium (DPAC,
\url{https://www.cosmos.esa.int/web/gaia/dpac/consortium}). Funding
for the DPAC has been provided by national institutions, in particular
the institutions participating in the {\it Gaia} Multilateral Agreement. 
Funding for the Sloan Digital Sky Survey IV has been provided by the Alfred P.\ Sloan Foundation, the U.S.\ Department of Energy Office of Science, and the Participating Institutions. SDSS acknowledges support and resources from the Center for High-Performance Computing at the University of Utah. The SDSS web site is \url{www.sdss.org}. SDSS is managed by the Astrophysical Research Consortium for the Participating Institutions of the SDSS Collaboration including the Brazilian Participation Group, the Carnegie Institution for Science, Carnegie Mellon University, the Chilean Participation Group, the French Participation Group, Harvard-Smithsonian Center for Astrophysics, Instituto de Astrofísica de Canarias, The Johns Hopkins University, Kavli Institute for the Physics and Mathematics of the Universe (IPMU) / University of Tokyo, Lawrence Berkeley National Laboratory, Leibniz Institut für Astrophysik Potsdam (AIP), Max-Planck-Institut für Astronomie (MPIA Heidelberg), Max-Planck-Institut für Astrophysik (MPA Garching), Max-Planck-Institut für Extraterrestrische Physik (MPE), National Astronomical Observatories of China, New Mexico State University, New York University, University of Notre Dame, Observatório Nacional / MCTI, The Ohio State University, Pennsylvania State University, Shanghai Astronomical Observatory, United Kingdom Participation Group, Universidad Nacional Autónoma de México, University of Arizona, University of Colorado Boulder, University of Oxford, University of Portsmouth, University of Utah, University of Virginia, University of Washington, University of Wisconsin, Vanderbilt University, and Yale University. 
Based on observations made with the NASA Galaxy Evolution Explorer. GALEX is operated for NASA by the California Institute of Technology under NASA contract NAS5-98034. 
The Pan-STARRS1 Surveys (PS1) and the PS1 public science archive have been made possible through contributions by the Institute for Astronomy, the University of Hawaii, the Pan-STARRS Project Office, the Max-Planck Society and its participating institutes, the Max Planck Institute for Astronomy, Heidelberg and the Max Planck Institute for Extraterrestrial Physics, Garching, The Johns Hopkins University, Durham University, the University of Edinburgh, the Queen's University Belfast, the Harvard-Smithsonian Center for Astrophysics, the Las Cumbres Observatory Global Telescope Network Incorporated, the National Central University of Taiwan, the Space Telescope Science Institute, the National Aeronautics and Space Administration under Grant No. NNX08AR22G issued through the Planetary Science Division of the NASA Science Mission Directorate, the National Science Foundation Grant No. AST-1238877, the University of Maryland, Eotvos Lorand University (ELTE), the Los Alamos National Laboratory, and the Gordon and Betty Moore Foundation. 
This publication makes use of data products from the Two Micron All Sky Survey, which is a joint project of the University of Massachusetts and the Infrared Processing and Analysis Center/California Institute of Technology, funded by the National Aeronautics and Space Administration and the National Science Foundation. 
This publication makes use of data products from the Wide-field Infrared Survey Explorer, which is a joint project of the University of California, Los Angeles, and the Jet Propulsion Laboratory/California Institute of Technology, funded by the National Aeronautics and Space Administration. 
\end{acknowledgements}
\bibliographystyle{aa}

%
\begin{appendix}
\renewcommand{\thefigure}{A.\arabic{figure}}
\renewcommand{\thetable}{A.\arabic{figure}}
\renewcommand{\theequation}{A.\arabic{equation}}
\begin{figure*}
\centering
\includegraphics[height=1\textwidth, angle=-90]{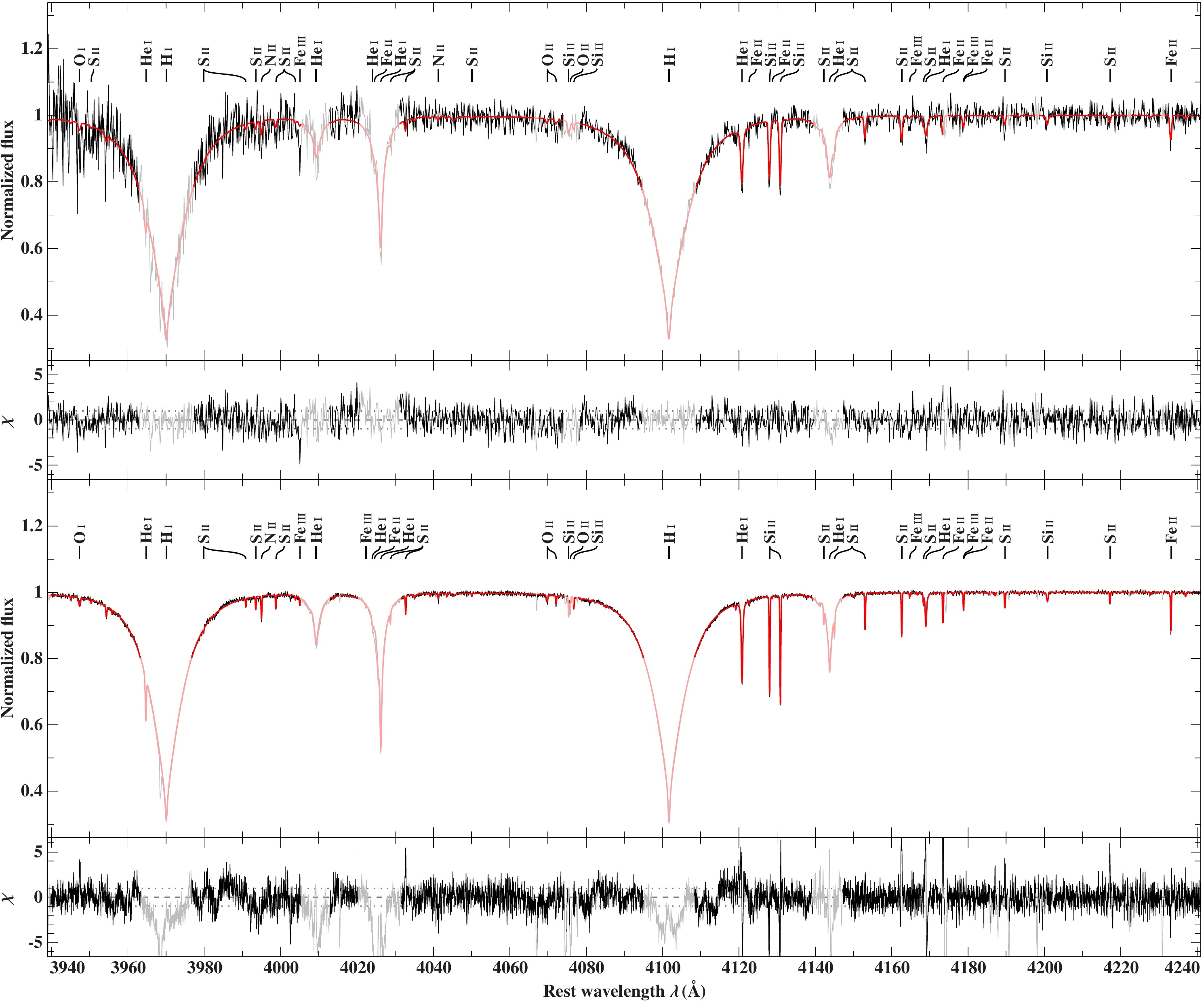}
\caption{Comparison of best-fitting model spectrum (red line) with normalized observed spectrum (black line) for HD\,137366 (\textit{left}, {\sc Feros}) and PG\,1610$+$062 (\textit{right}, ESI). Light colors mark regions that have been excluded from fitting (e.g., due to the presence of features that are not properly included in our models). For the sake of clarity, only the strongest of the lines used in the analysis are labeled. Residuals $\chi$ are shown as well. Telluric correction is performed via interpolation within the pre-calculated grid of transmission spectra presented by \citetads{2014A&A...568A...9M}. Although the atmospheric conditions used in this spectral library are tailored to Cerro Paranal, the two free parameters of airmass and precipitable water vapor content are enough to ensure a decent representation of many telluric features for different observing sites and weather conditions. Regions where telluric features are not properly reproduced by this approach have been excluded as well.}
\label{fig:spectra_1}
\end{figure*}
\begin{figure*}
\centering
\includegraphics[height=1\textwidth, angle=-90]{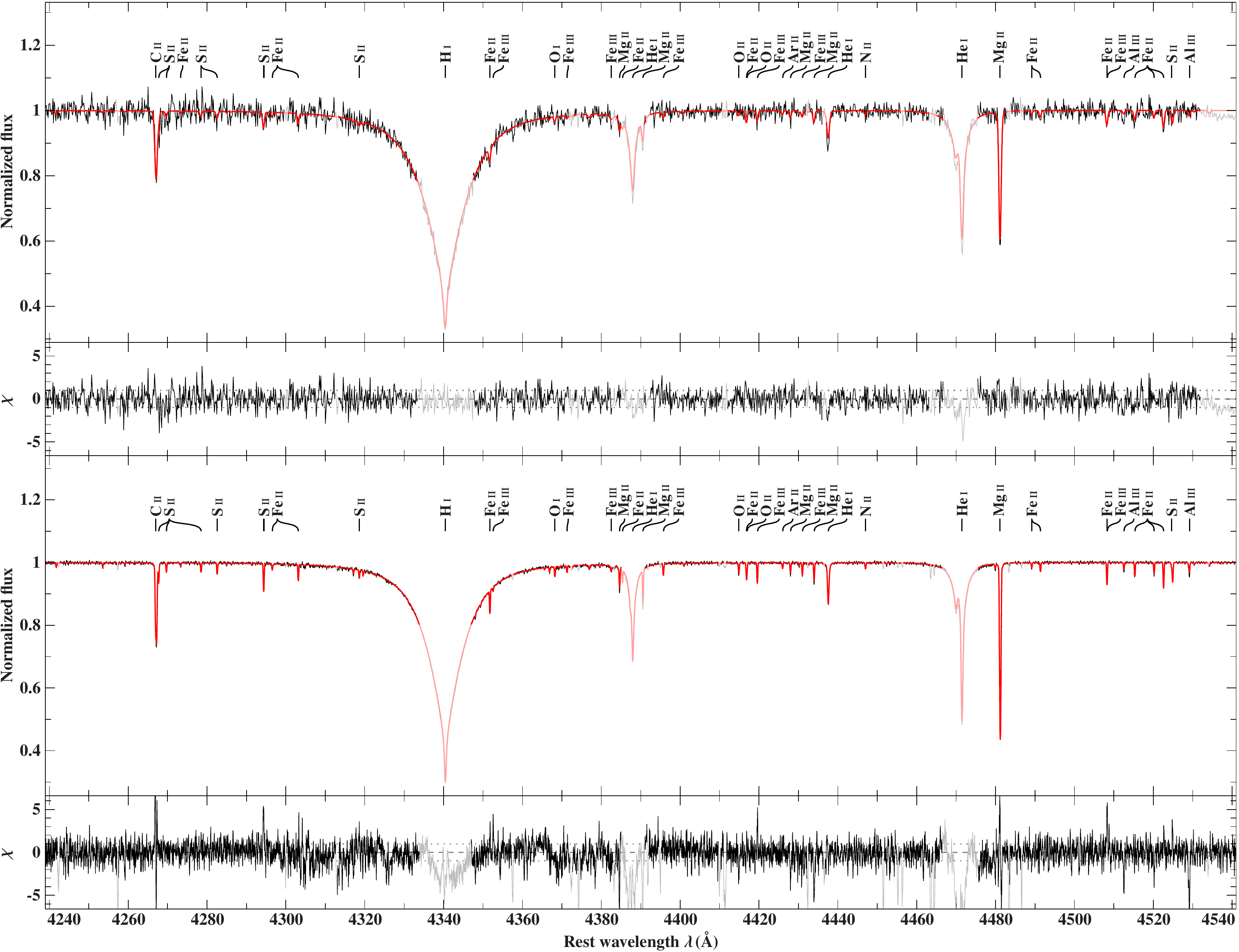}
\caption{Same as Fig.~\ref{fig:spectra_1}.}
\label{fig:spectra_2}
\end{figure*}
\begin{figure*}
\centering
\includegraphics[height=1\textwidth, angle=-90]{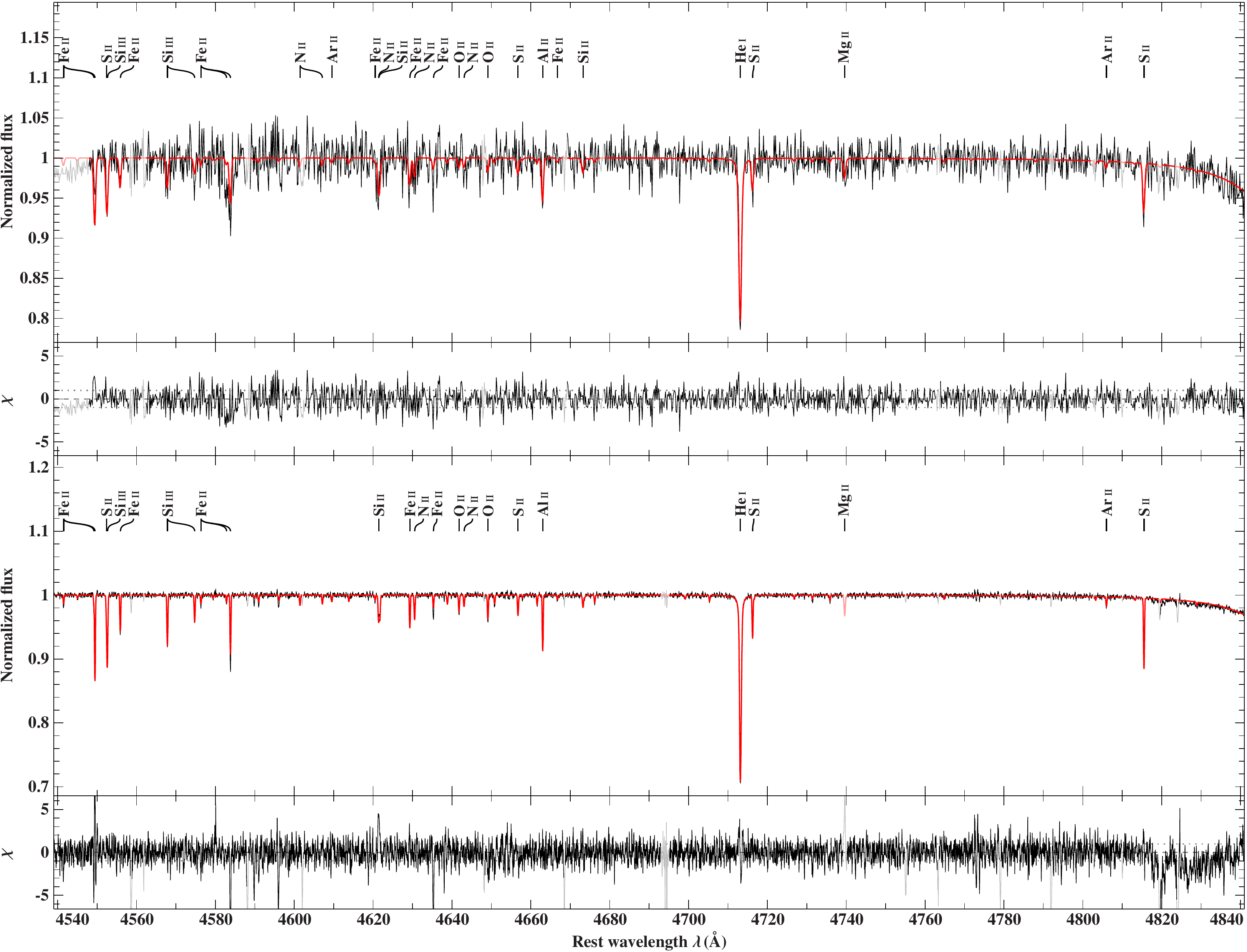}
\caption{Same as Fig.~\ref{fig:spectra_1}.}
\label{fig:spectra_3}
\end{figure*}
\begin{figure*}
\centering
\includegraphics[height=1\textwidth, angle=-90]{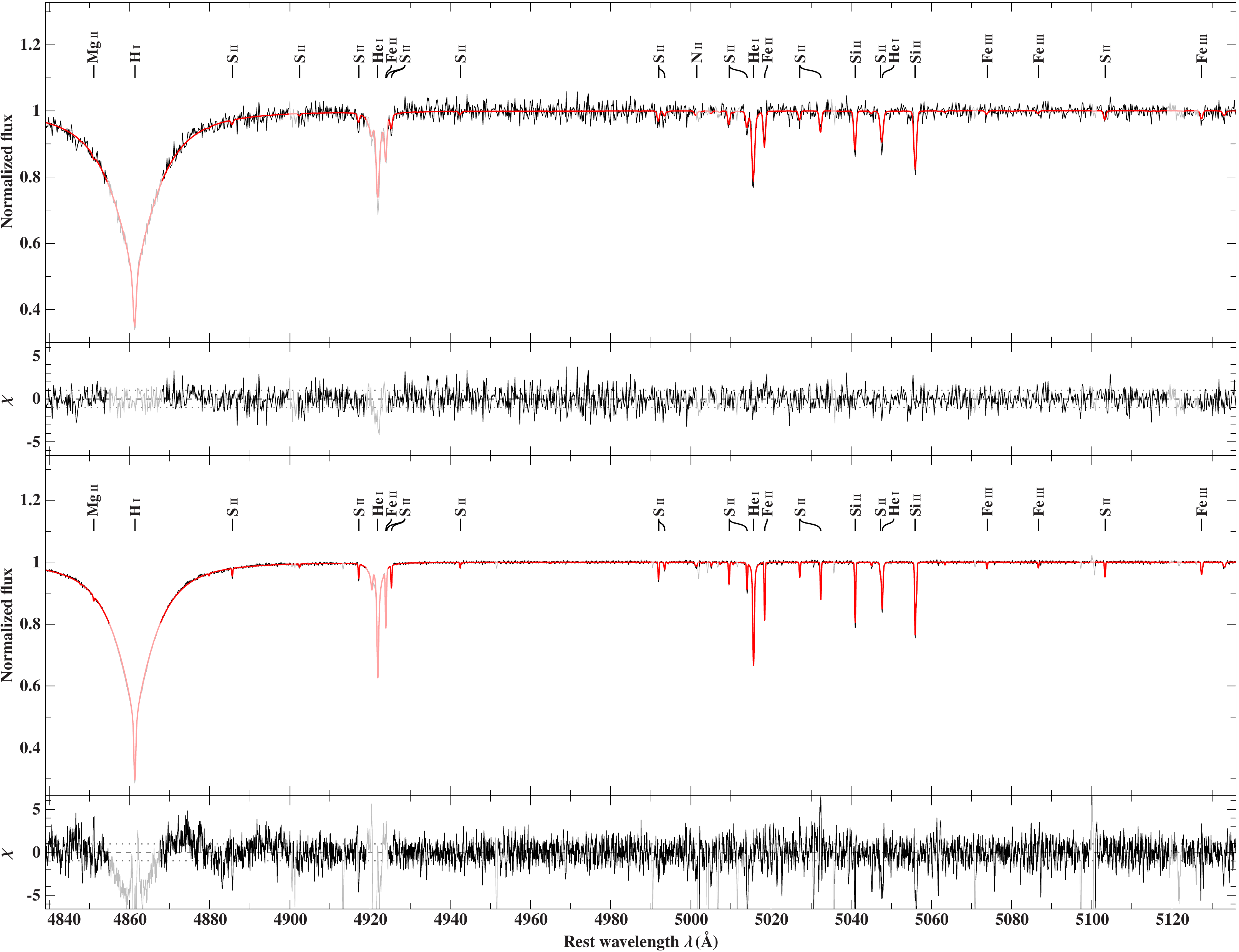}
\caption{Same as Fig.~\ref{fig:spectra_1}.}
\label{fig:spectra_4}
\end{figure*}
\begin{figure*}
\centering
\includegraphics[height=1\textwidth, angle=-90]{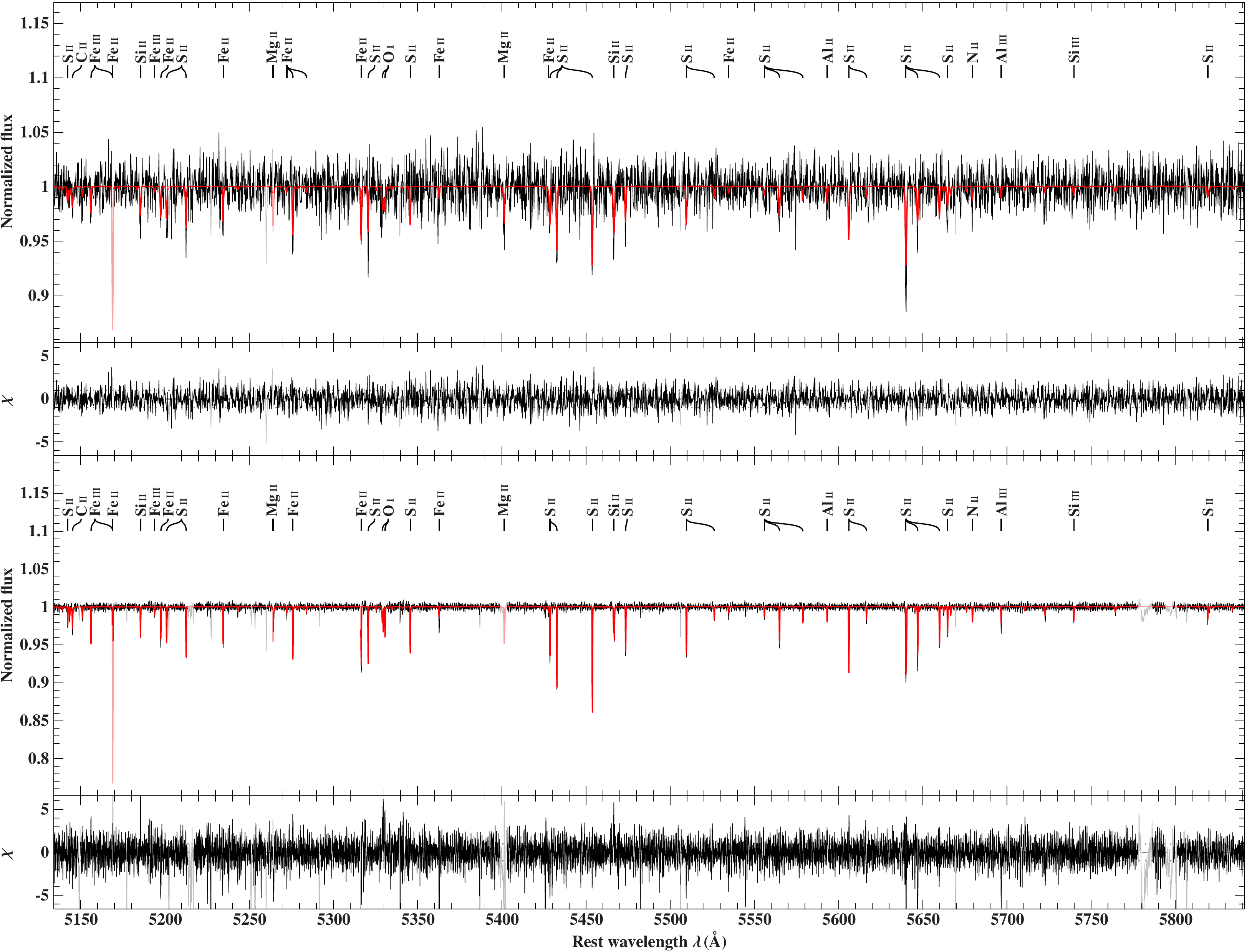}
\caption{Same as Fig.~\ref{fig:spectra_1}.}
\label{fig:spectra_5}
\end{figure*}
\begin{figure*}
\centering
\includegraphics[height=1\textwidth, angle=-90]{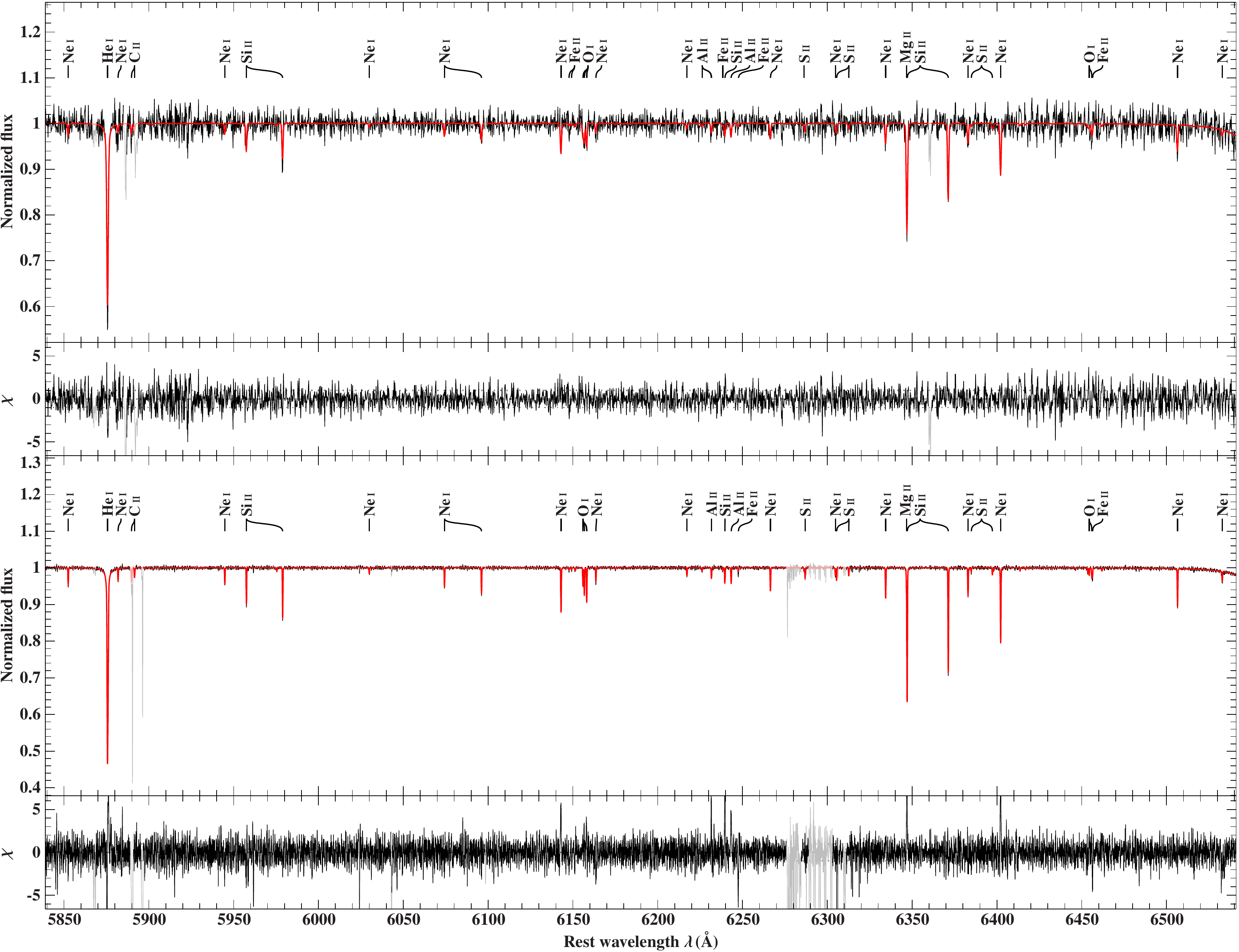}
\caption{Same as Fig.~\ref{fig:spectra_1}.}
\label{fig:spectra_6}
\end{figure*}
\begin{figure*}
\centering
\includegraphics[height=1\textwidth, angle=-90]{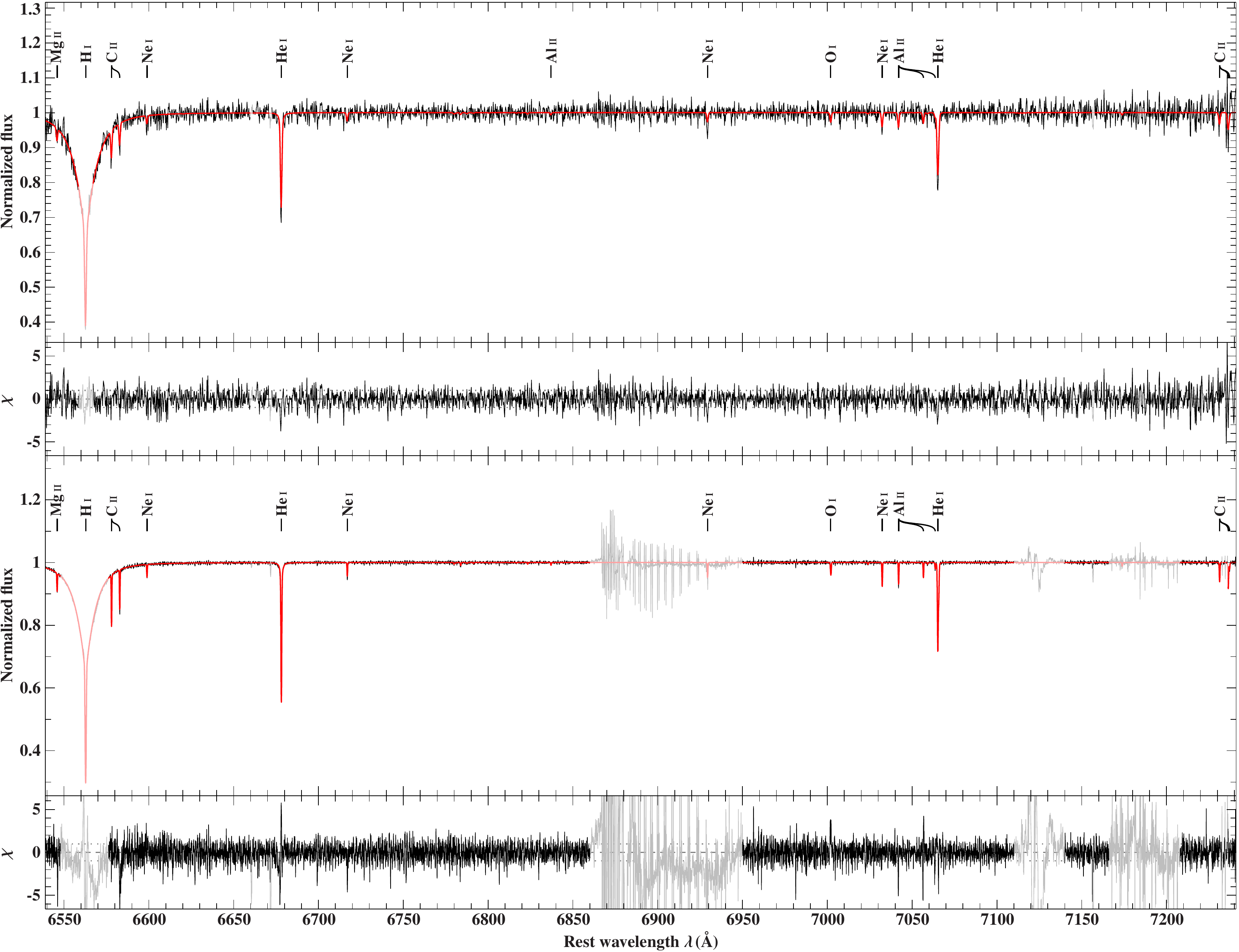}
\caption{Same as Fig.~\ref{fig:spectra_1}.}
\label{fig:spectra_7}
\end{figure*}
\begin{figure*}
\centering
\includegraphics[height=1\textwidth, angle=-90]{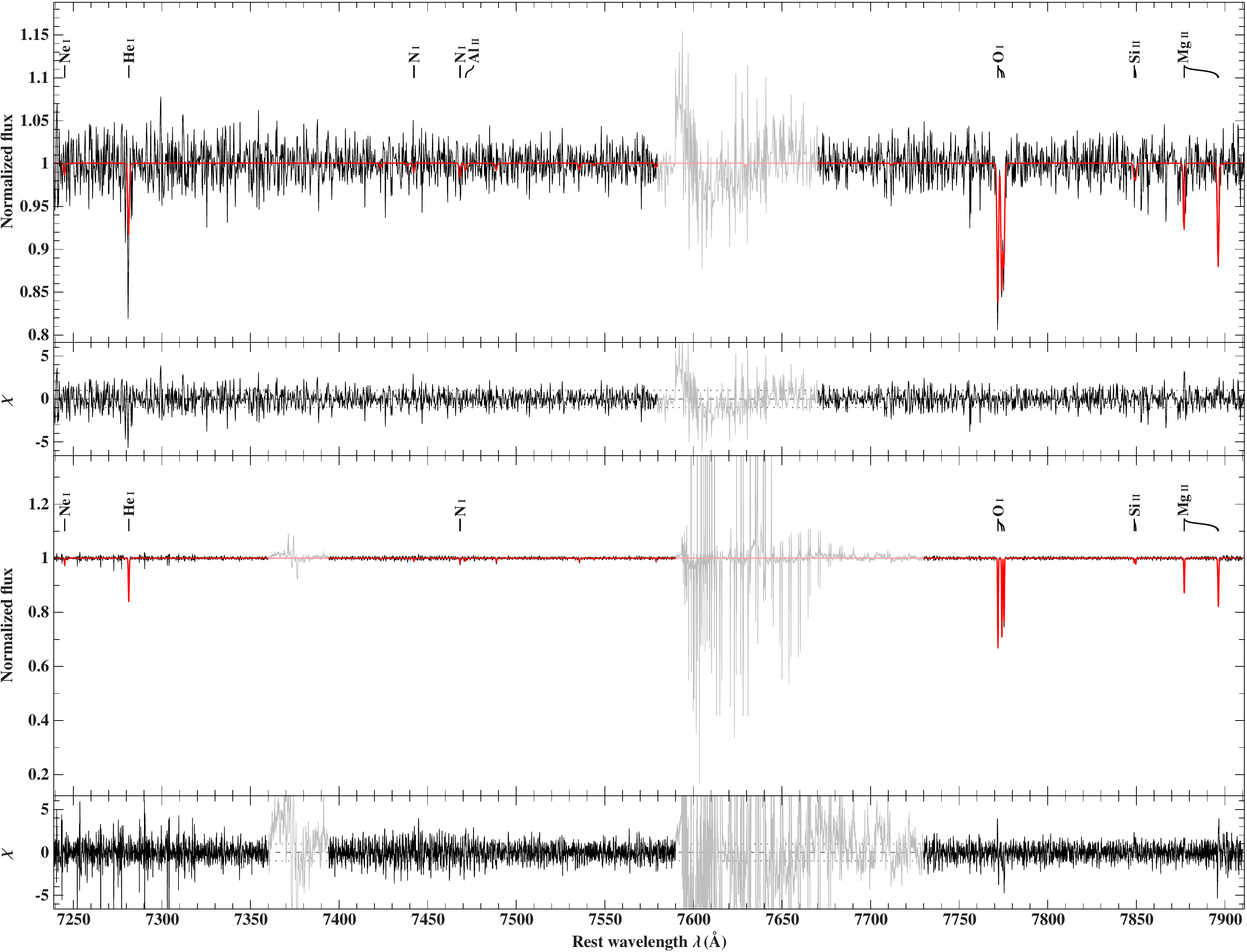}
\caption{Same as Fig.~\ref{fig:spectra_1}.}
\label{fig:spectra_8}
\end{figure*}
\begin{figure*}
\centering
\includegraphics[height=1\textwidth, angle=-90]{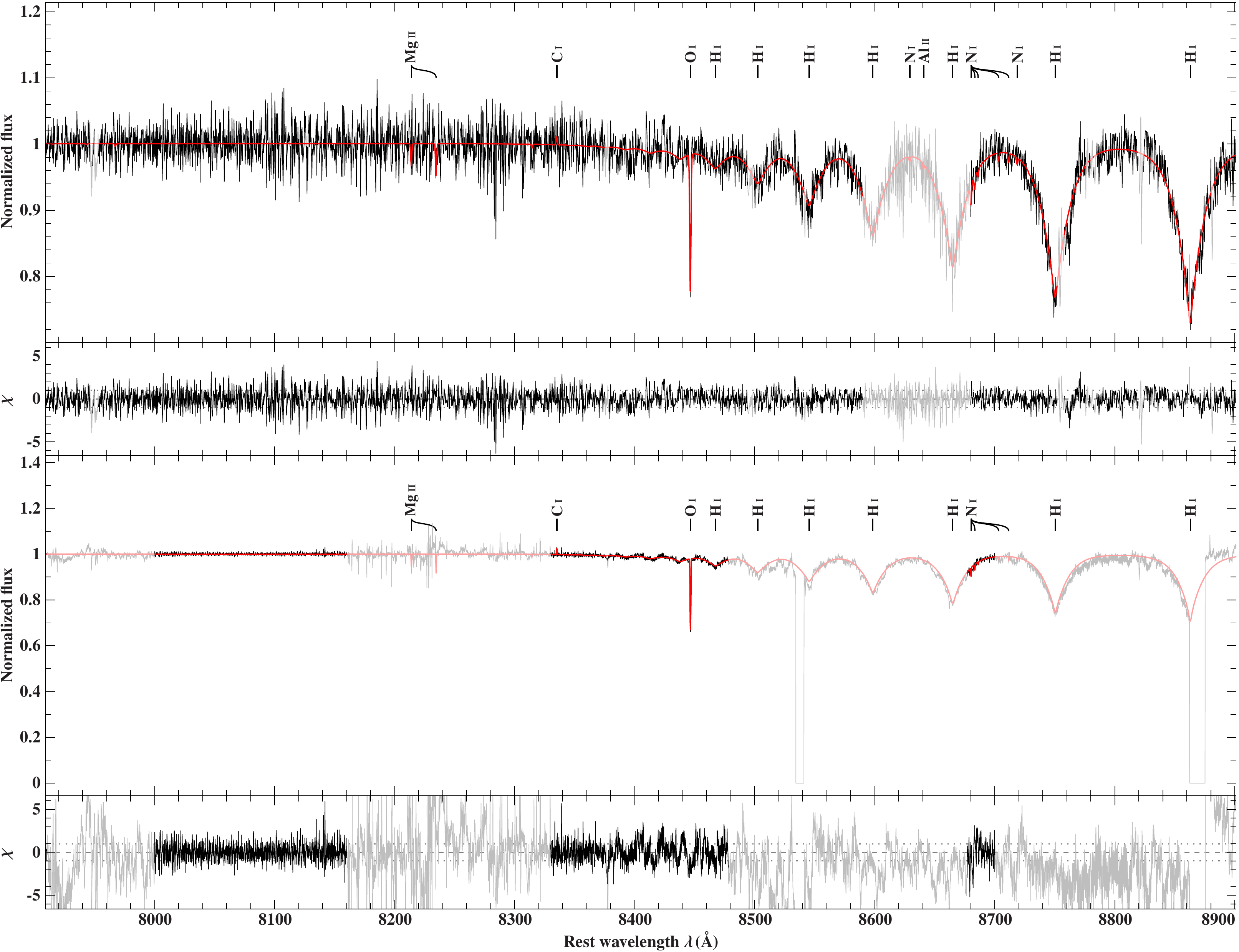}
\caption{Same as Fig.~\ref{fig:spectra_1}.}
\label{fig:spectra_9}
\end{figure*}
\begin{figure*}
\includegraphics[height=0.57\textwidth]{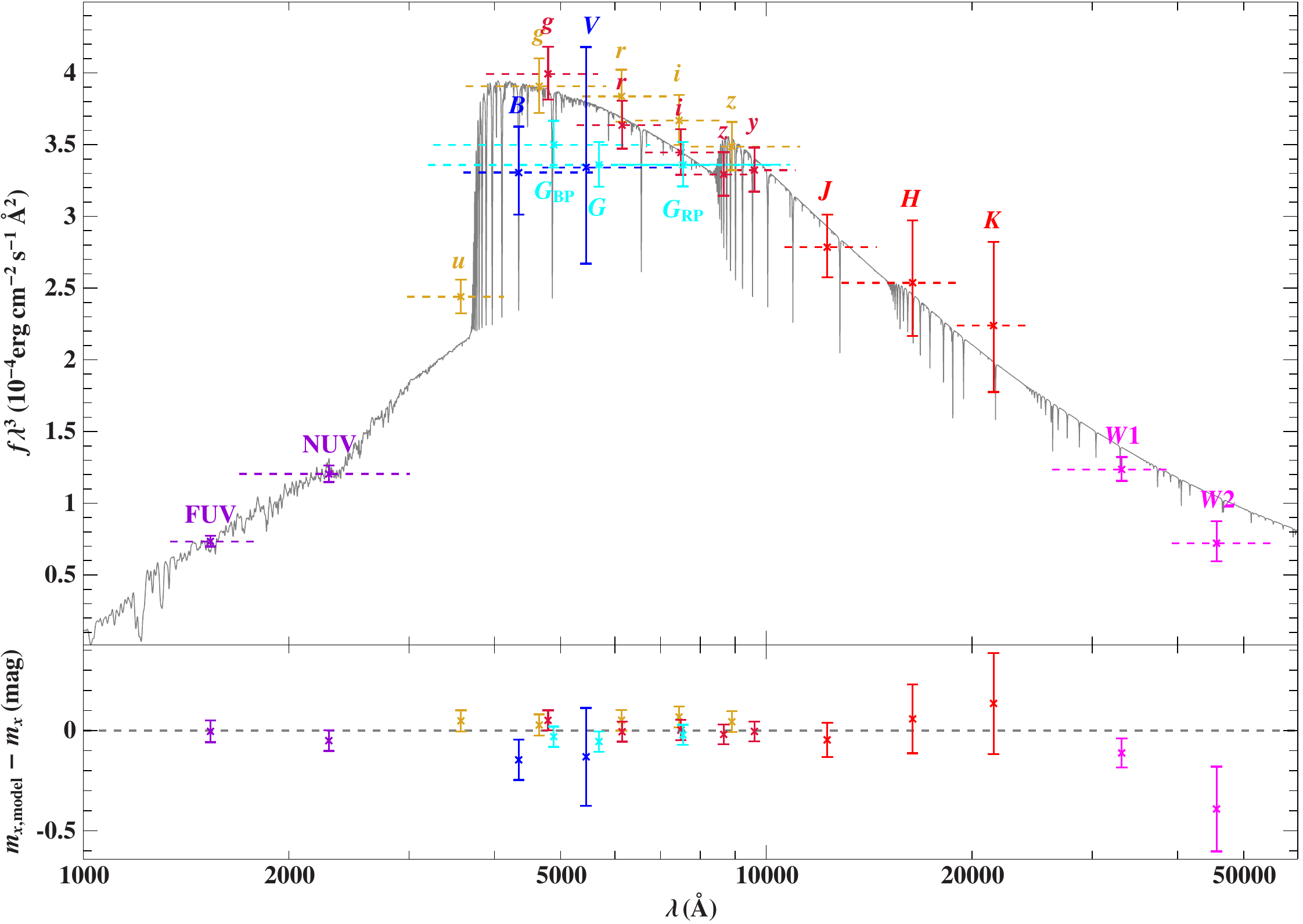}
\includegraphics[height=0.57\textwidth]{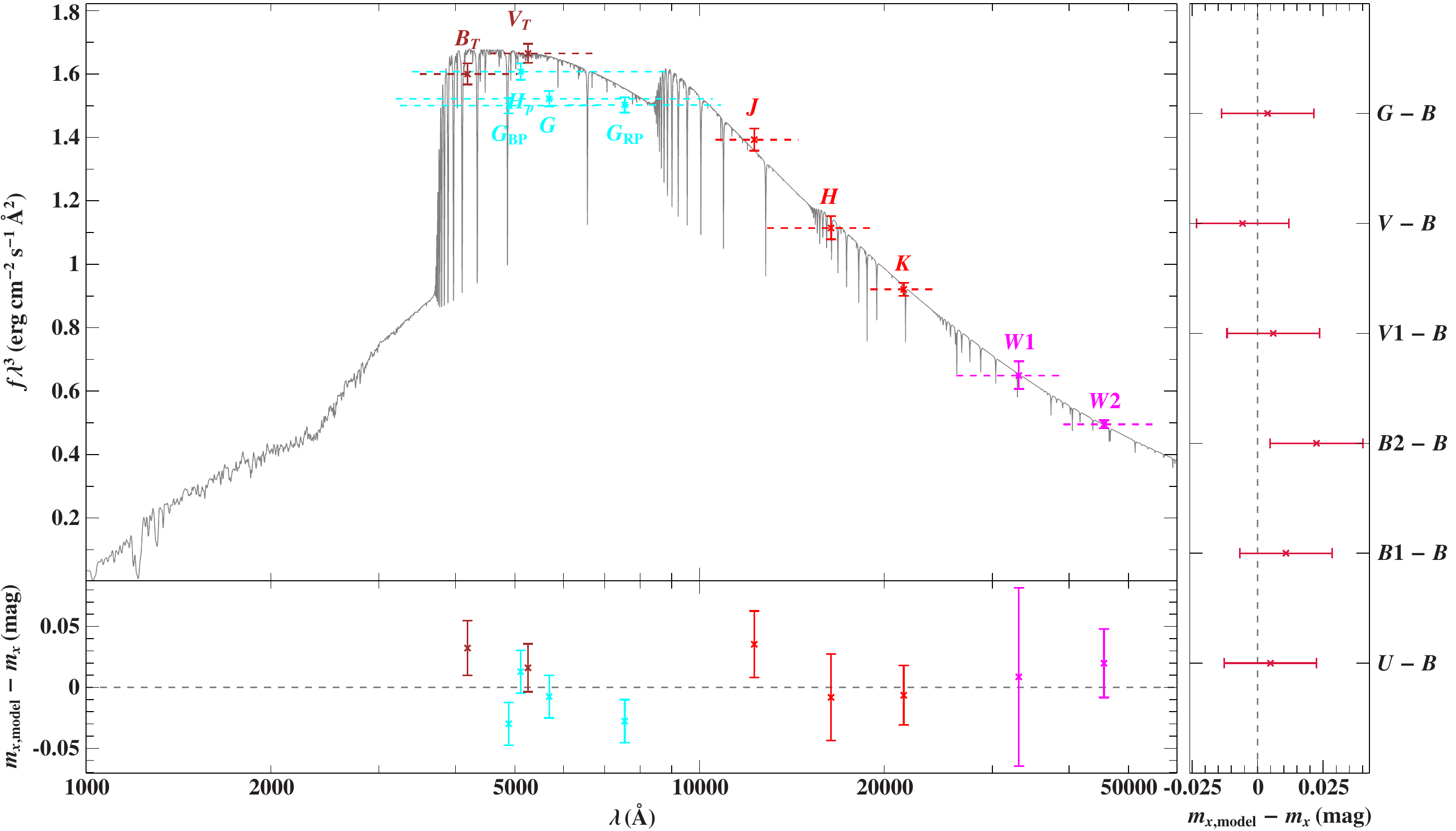}
\caption{Comparison of synthetic and observed photometry for PG\,1610$+$062 (\textit{top figure}) and HD\,137366 (\textit{bottom figure}). The main panels show the spectral energy distributions. The colored data points are filter-averaged fluxes which were converted from observed magnitudes (the respective filter widths are indicated by the dashed horizontal lines), while the gray solid line represents the best-fitting model (degraded to a spectral resolution of 6\,{\tiny\AA}). The residual panels at the bottom and on the side show the differences between synthetic and observed magnitudes and colors, respectively. The extinction law by \citetads{1999PASP..111...63F} with the color excess as free parameter was used to account for interstellar reddening. The photometric systems are color-coded as follows: violet: GALEX (DR5, \citeads{2011Ap&SS.335..161B}, corrected using the description given in \citeads{2014MNRAS.438.3111C}); gold: SDSS (DR9, \citeads{2012ApJS..203...21A}); blue: Johnson-Cousins (APASS DR9, \citeads{2015AAS...22533616H}); cyan: {\it Gaia} \citepads{2018A&A...616A...4E}, {\sc Hipparcos} \citepads{2007ASSL..350.....V}; brown: {\sc Tycho} \citepads{2007ASSL..350.....V}; crimson: Pan-STARRS \citepads{2017yCat.2349....0C}, Geneva \citepads{1988csmg.book.....R}; red: 2MASS \citepads{2006AJ....131.1163S}; magenta: WISE \citepads{2014yCat.2328....0C}.}
\label{fig:photometry}
\end{figure*}
\clearpage
\begin{table}
\centering
\caption{Observed oscillation parameters of PG\,1610$+$062 derived from the two ATLAS light curves.}
\label{table:oscillation_params}
\footnotesize
\begin{tabular}{lr}
\hline\hline
Parameter & Value \\
\hline
Frequency $\nu_{\textnormal{osc}}$ & $0.23052\pm0.00022$\,d$^{-1}$ \\
Period $P_{\textnormal{osc}}$ & $4.338\pm0.005$\,d \\
Reference epoch $T_{\textnormal{ref}}$ (fixed) & $57\,230.0$\,MJD \\
Phase $\phi_{\textnormal{ref}}$ at epoch $T_{\textnormal{ref}}$ & $0.30\pm0.08$ \\
$o$ mean magnitude & $15.746\pm0.008$\,mag \\
$o$ semiamplitude & $21\pm11$\,mmag \\
$c$ mean magnitude & $15.457\pm0.006$\,mag \\
$c$ semiamplitude & $35\pm7$\,mmag \\
\hline
\end{tabular}
\tablefoot{The given uncertainties are single-parameter 99\% confidence intervals based on the $\chi^2$ statistics around the best fit with $\chi^2_\textnormal{reduced} \approx 1.5$.}
\end{table}
\begin{figure}
\includegraphics[width=0.49\textwidth]{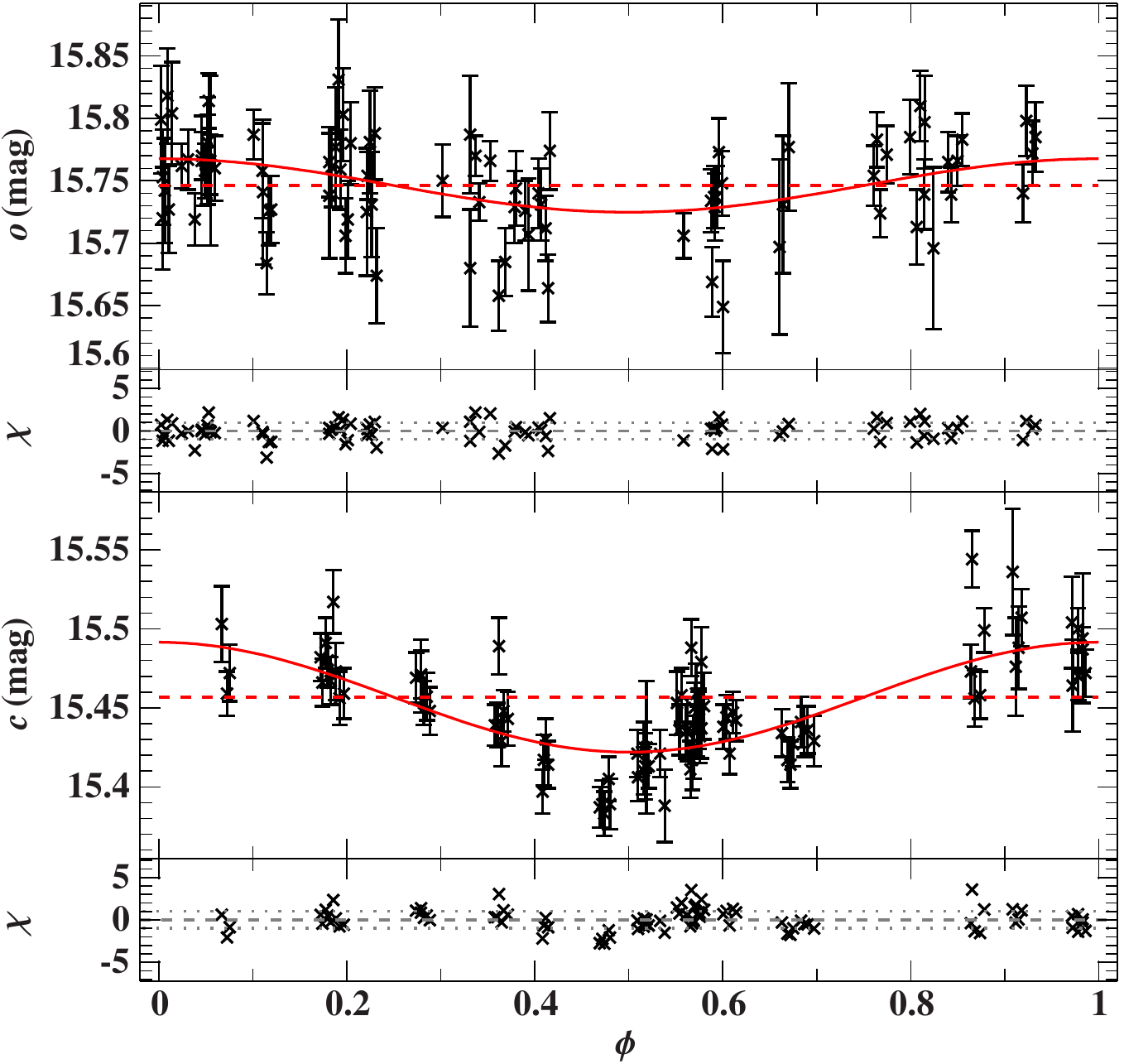}
\caption{Phased ATLAS light curves for PG\,1610$+$062: the measurements are represented by black crosses with error bars while the best-fitting model (see Table~\ref{table:oscillation_params}) is indicated by the red solid curve. The red dashed line indicates the derived mean magnitude. Residuals $\chi$ are shown as well.}
\label{fig:phased_lightcurves}
\end{figure}
\begin{figure}
\includegraphics[width=0.49\textwidth]{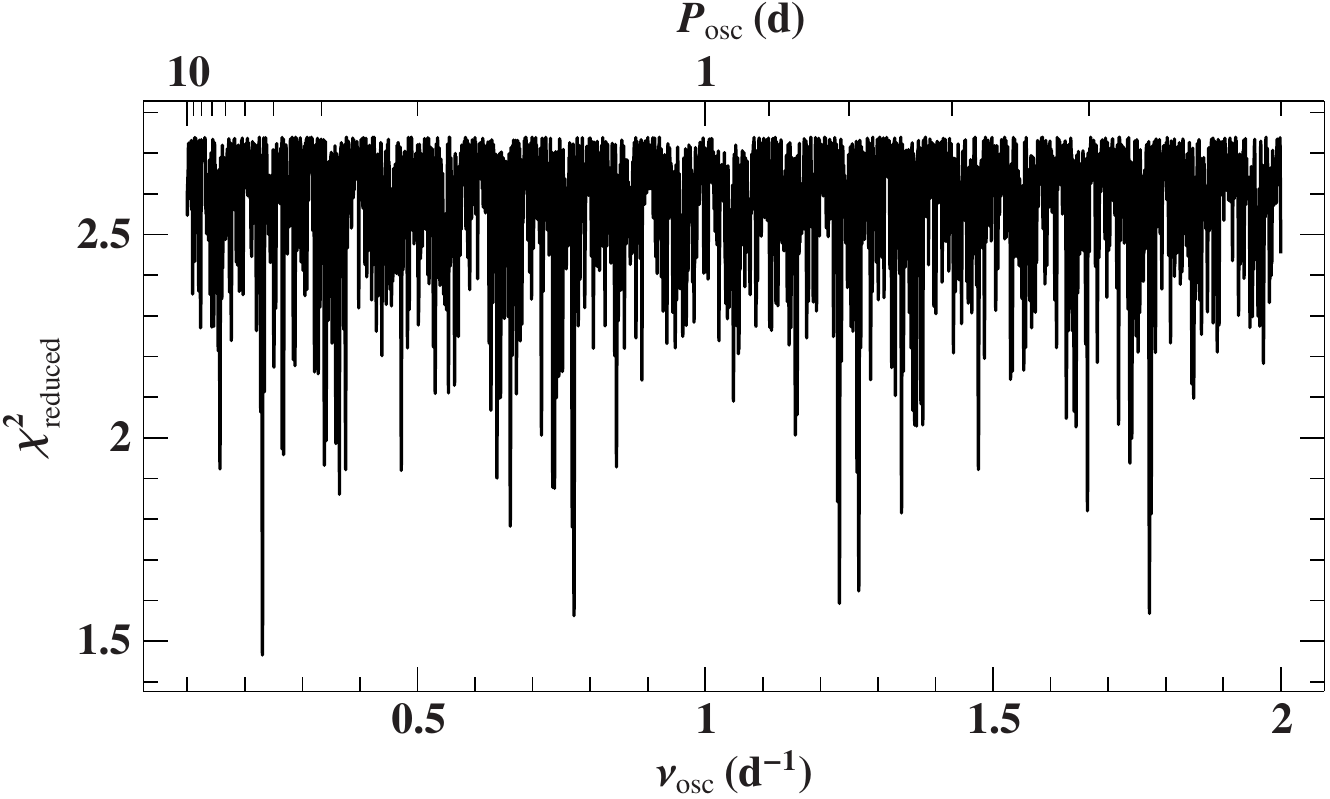}
\caption{The $\chi^2$ landscape (``periodogram''), which results from fitting the ATLAS light curves with the model given in Eq.~(\ref{eq:cosine_fit}), as a function of the oscillation frequency $\nu_{\textnormal{osc}}$, which is sampled in steps of $0.01/686.25$\,d$^{-1}$ to ensure that phase shifts are always less than $0.01$.}
\label{fig:periodogram}
\end{figure}
\noindent After removing a few obvious outliers, the available ATLAS light curves (which consist of 90 measurements in the cyan ($c$) band spread over $653.29$\,days and 88 data points in the orange ($o$) band spread over $686.25$\,days) were fitted with a cosine function of the form
\begin{equation}%
\textnormal{mag}_j(t) = \overline{\textnormal{mag}}_j + A_{j} \cos\left(2\pi\left[(t-T_{\textnormal{ref}}) \nu_\textnormal{osc}+\phi_{\textnormal{ref}}\right] \right) \,.
\label{eq:cosine_fit}
\end{equation}
The time-dependent magnitude $\textnormal{mag}_j(t)$ is thus parameterized by a mean magnitude $\overline{\textnormal{mag}}_j$, an oscillation semiamplitude $A_{j}$, and an oscillation frequency $\nu_\textnormal{osc}$. The parameter $\phi_{\textnormal{ref}}$ is the phase at the fixed reference epoch $T_{\textnormal{ref}}$. The index $j \in \{c, o\}$ refers to the two passbands. The best-fitting parameters are listed in Table~\ref{table:oscillation_params} and the corresponding phased light curves are shown in Fig.~\ref{fig:phased_lightcurves}. The oscillation period reported by \citetads{2018AJ....156..241H}, $4.336721$\,days, is nicely confirmed here. However, due to the very scarce sampling, alias frequencies at $1-\nu_{\textnormal{osc}}$, $1+\nu_{\textnormal{osc}}$, and $2-\nu_{\textnormal{osc}}$ are almost as likely as $\nu_{\textnormal{osc}}$ itself (see Fig.~\ref{fig:periodogram}). Without better data coverage, it remains unclear which of them is actually the true one.
\begin{figure}
\centering
\includegraphics[width=0.49\textwidth]{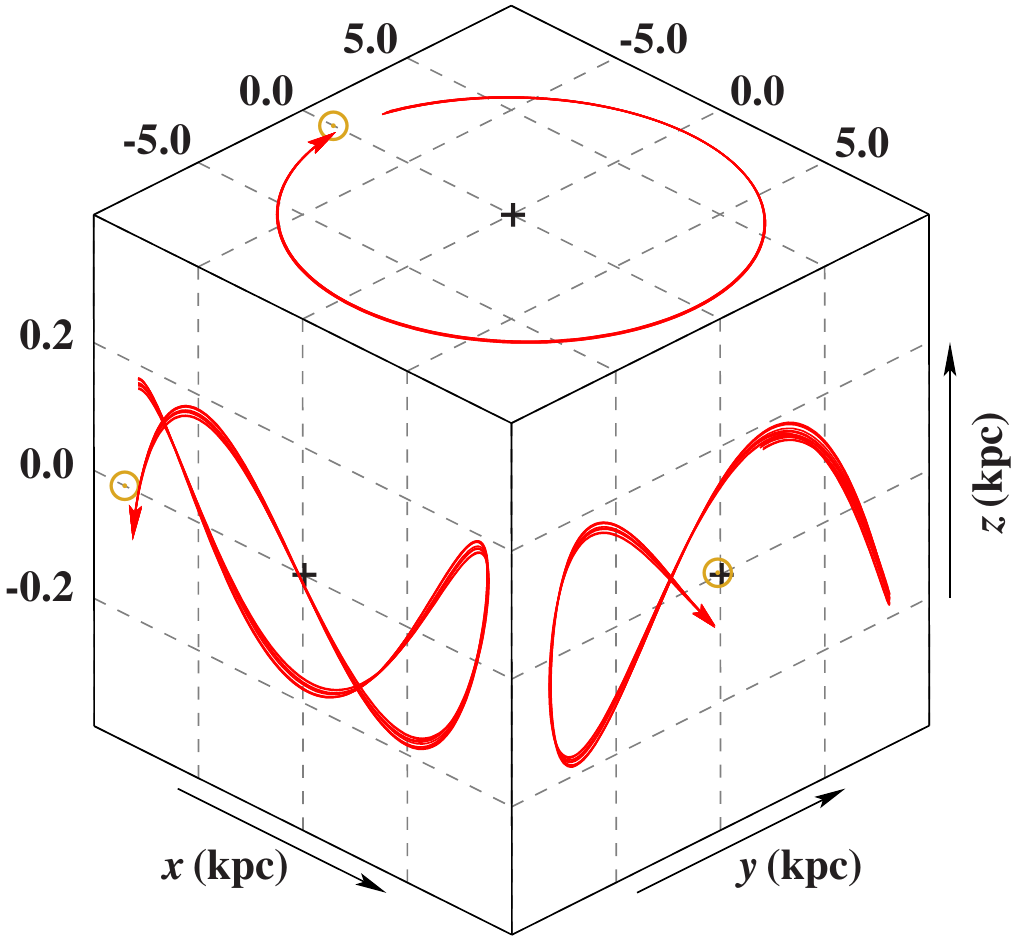}
\caption{Same as Fig.~\ref{fig:orbits_PG1610}, but for HD\,137366. Orbits were computed back in time for 200\,Myr. The star behaves like a typical solar-neighborhood object.
}\label{fig:orbits_HD137366}
\end{figure}
\end{appendix}
\end{document}